\documentclass[
reprint,
superscriptaddress,
amsmath,
amssymb,
aps,
pra,
floatfix,
]{revtex4-2}

\usepackage{graphicx}
\usepackage{dcolumn}
\usepackage{bm}
\usepackage{soul}
\usepackage{braket}
\usepackage[dvipsnames]{xcolor}
\usepackage{mathtools}
\usepackage{array}
\usepackage{enumerate}   
\usepackage{siunitx}
\usepackage{lipsum}
\usepackage[normalem]{ulem}
\usepackage[colorlinks=true,urlcolor=blue,citecolor=blue,linkcolor=blue]{hyperref}

\DeclareMathOperator{\pump}{\mathrm{p}}
\DeclareMathOperator{\signal}{\mathrm{s}}
\DeclareMathOperator{\idler}{\mathrm{i}}

\begin{document}

\preprint{APS/123-QED}

\title{Engineering of maximally entangled orbital angular momentum states \\ via path identity}

\author{Richard Bernecker}
 \email{richard.bernecker@uni-jena.de}
\affiliation{
 Theoretisch-Physikalisches Institut, Friedrich-Schiller-Universität Jena, Max-Wien-Platz 1, D-07743 Jena, Germany}
\affiliation{
 Helmholtz-Institut Jena, Fröbelstieg 3, D-07743 Jena, Germany 
}
 
\author{Baghdasar Baghdasaryan}
\affiliation{
Institut für Angewandte Physik, Friedrich-Schiller-Universität Jena, Albert-Einstein-Str. 6, D-07745 Jena, Germany
}

\author{Stephan Fritzsche}
\affiliation{
 Theoretisch-Physikalisches Institut, Friedrich-Schiller-Universität Jena, Max-Wien-Platz 1, D-07743 Jena, Germany}
\affiliation{
 Helmholtz-Institut Jena, Fröbelstieg 3, D-07743 Jena, Germany 
}

\date{\today}

\begin{abstract} 
Cutting-edge quantum technologies lean on sources of high-dimensional entangled states (HDES) that reliably prepare high-fidelity target states. The idea to overlap photon paths from distinct but indistinguishable sources was recently introduced for the creation of HDES, known as entanglement by path identity. In this regard, the use of orbital angular momentum (OAM) modes is promising, as they offer a high-dimensional and discrete Hilbert space to encode information. While entanglement by path identity with OAM has been verified experimentally, a detailed investigation of how the OAM distribution of photon pairs can be engineered to maximize the entanglement is lacking. We address this gap and identify an optimal dimensionality for maximally entangled states (MESs) when the spatial engineering of the pump beam and the path identity approach are combined. Our theoretical study reveals notable limitations for the fidelity of high-dimensional target states. We also establish the equivalence of entangled biphoton states pumped by a spatially engineered beam and generated via path identity. These findings constitute a valuable step toward the optimized preparation of MESs in high dimensions.
\end{abstract}

\maketitle

\section{Introduction}
Entangled photon pairs are a crucial resource to build the future quantum internet and underpin secure quantum communication and distributed quantum computing \cite{sidhu2021advances,zhong2020quantum, azuma2023quantum,chen2021review, caleffi2024distributed}. Especially the entanglement between $d>2$-dimensional quantum systems, called \textit{qudits}, offers unique advantages compared to the use of two-dimensional qubits. These include enhanced capabilities with superdense coding, increased noise resistance, and higher error tolerance in quantum information processing applications \cite{wang2020qudits,cozzolino2019high}. 

Photon-pair sources are usually based on the probabilistic interaction of an intense pump laser with nonlinear materials, as in spontaneous parametric down-conversion (SPDC) \cite{anwar2021entangled}. Remarkably, photonic qudits can be encoded in a wide range of degrees of freedom (DOFs), including photon number \cite{bimbard2010quantum,thekkadath2020quantum}, polarization \cite{bogdanov2004qutrit,fedorov2011entanglement,sekga2023high}, spatial modes \cite{krenn2017orbital,forbes2019quantum}, temporal and frequency modes \cite{kues2017chip,martin2017quantifying,serino2024orchestrating}, or time bins  \cite{donohue2013coherent,grassani2015micrometer}. 

Among these DOFs, transverse spatial modes of photons that carry OAM stand out as a prime candidate to realize bipartite HDES. \cite{erhard2020advances,zhang2024entanglement}. Their suitability for long-distance communication, both in fiber \cite{cao2020distribution} and in free-space \cite{krenn2015twisted,sit2017high}, alongside efficient generation and progressively improving measurement techniques \cite{krenn2014generation,fickler2016quantum,bouchard2018measuring,hiekkamaki2019near,karan2025broadband}, positions them as a scalable and practical platform for advanced quantum communication protocols \cite{sun2024experimental,zhao2025high,scarfe2025high}.

While SPDC naturally generates spatially entangled photon pairs based on the conservation of OAM \cite{mair2001entanglement,li2019generation}, a key limitation exists: the broad but non-uniform OAM distribution of the down-converted pair \cite{di2010measurement,yao2011angular,pors2011high} prevents the formation of MESs \cite{horodecki2009quantum,nielsen2010quantum}. These states are characterized by uniformly distributed, non-separable OAM modes, where the amplitudes of $d$ constituent modes are precisely $1/\sqrt{d}$, and all other OAM modes have zero amplitude.

Various methods have been studied to tune the OAM distribution of the photon pair, which span from modification of the spatial spectrum of the pump beam \cite{liu2020increasing,baghdasaryan2020enhanced,torres2003preparation,boucher2021engineering} to a deliberate control over the nonlinearity of crystals \cite{rozenberg2022inverse,bernecker24high,xu2021manipulating}. Although equally probable and non-separable OAM modes are prepared, the current methods are restricted to a projection \textit{subspace} of the full Hilbert space \cite{liu2018coherent,kovlakov2018quantum,bornman2021optimal}. The full state is not maximally entangled, since OAM modes outside this defined subspace remain undetected. One direct consequence is a reduced detection rate, which can significantly limit the achievable key rate in quantum key distribution \cite{zhang2021security}.

A recent idea to customize photon pairs is entanglement by path identity \cite{krenn2017entanglement}. Here, the experimental setup ensures a coherent overlap of photon paths, placing the pair in a superposition of multiple indistinguishable origins. A proof-of-concept experiment \cite{kysela2020path} demonstrated the versatile utility of path identity with OAM modes, but still faces the limitation of non-maximal HDES preparation outside a specific subspace.

To overcome this drawback, we present a comprehensive investigation that merges the advantages of pump beam shaping with path-identity principles. Our method enables the creation of high-fidelity maximally entangled OAM states. Interestingly, our study uncovers fundamental fidelity bounds for target states. We also formulate conditions under which the path-identity approach and pump-beam engineering are equivalent. 

Finally, we demonstrate that the optimal dimensionality $d$ for preparing true MESs in path-identity setups with $n$  nonlinear crystals is $d=2n$. We provide an example by designing setups with $n=2$ coherently pumped crystals to generate a maximally entangled ququarts ($d=4$). Those resulting states are generalized four-dimensional Bell states, which are beneficial for high-dimensional entanglement swapping protocols \cite{zhang2017simultaneous, zangi2025swapped, baghdasaryan2025efficient} or to beat the channel capacity limit for superdense coding \cite{hu2018beating}. Our findings are also readily scalable to setups with additional crystals, opening new avenues for the efficient generation of high-dimensional MESs.

\section{Theoretical Foundations}

SPDC is a three-wave mixing process where an intense pump beam interacts with a medium exhibiting a second-order nonlinearity $\chi^{(2)}$. The interaction generates correlated photon pairs, also known as signal and idler. The energy and momentum are strictly conserved in the SPDC process. Because the intensity of the pump beam far exceeds the low generation rate of the photon pairs, the pump beam is typically modeled as a classical field.

\subsection{Biphoton state in OAM basis}
\label{Photon pair generation}
To accurately grasp experiments involving OAM mode entanglement, we first derive the quantum state of the photon pair, the \textit{biphoton state}, using well-known approximations \cite{walborn2010spatial,fabre2020modes,baghdasaryan2022maximizing,bernecker2023spatial,baghdasaryan2022generalized}.

Specifically within the paraxial approximation, the longitudinal wave vector component $k_z$ is much larger than the transverse momentum $\bm{q}$,  assuming the photons propagate quasi-collinearly along the $z$-axis. Using $k_z \gg \bm{q}$, allows us to write  $k_z(\bm{q},\omega) \approx k(\omega) - |\bm{q}|^2 / 2k(\omega)$. The longitudinal component is completely described in terms of the transverse momentum vector $\bm{q}$ and the wave number $k = \frac{n \omega}{c}$. Within this regime, the frequency $\omega$ is referred to as the \textit{spectral}, and $\bm{q}$ as the \textit{spatial} DOF. 

The longitudinal wave vector mismatch is expressed as $\Delta k_z(\bm{q}_{\signal},\bm{q}_{\signal}) = k_{\pump,z}(\bm{q}_{\signal}+\bm{q}_{\signal}) - k_{\signal,z}(\bm{q}_{\signal}) - k_{\idler,z}(\bm{q}_{\idler})$, where the subscripts $\pump, \signal,$ and $\idler$ denote the pump, signal, and idler, respectively. The conservation of transverse momentum $\bm{q}_{\pump} = \bm{q}_{\signal} + \bm{q}_{\idler}$ is enforced if the transverse dimension of the crystal (typically $\sim$ cm) is considerably larger than the size of the pump beam ($\sim \mu$m) \cite{
anwar2018direct,baghdasaryan2021justifying}.

We also employ the narrowband approximation, wherein the spectral bandwidths of the photons are assumed to be sufficiently small. In this regime, all frequency-dependent quantities can be evaluated at their central values, which satisfy the energy-conservation $\omega_{\pump} = \omega_{\signal} + \omega_{\idler}$. The narrowband regime is experimentally enforced using frequency filters \cite{van2020optimizing}. Under the given assumptions, the spatial biphoton state can be written as 
\begin{equation}
     \ket{\psi} = \mathcal{N}  \iint d\bm{q}_{\signal} \; d\bm{q}_{\idler} \; \Phi(\bm{q}_{\signal},\bm{q}_{\idler}) \Bigl( \hat{a}_{\signal}^{\dagger}(\bm{q}_{\signal}) \otimes \hat{a} _{\idler}^{\dagger}(\bm{q}_{\idler}) \Bigl)  \ket{\text{vac}}
     \label{GeneralBiphotonstate}   
\end{equation}
where $\mathcal{N}$ is a normalization constant and $\hat{a}^{\dagger}$ the creation operator for the signal and idler mode. The mode function $\Phi$ encodes the spatial correlations between the signal and idler photons via
\begin{equation}
    \Phi(\bm{q}_{\signal},\bm{q}_{\idler}) =  V_{\pump}(\bm{q}_{\signal}+\bm{q}_{\idler}) \int_{-\frac{L}{2}}^{\frac{L}{2}} dz \;  \chi^{(2)}(z) \; \mathrm{e}^{ i \Delta k_z(\bm{q}_{\signal},\bm{q}_{\signal}) z}.
\end{equation}
The mode function consists of two principal constituents: the transverse spatial distribution of the pump beam $V_{\pump}$, and the phase-matching function. The latter is an integral whose parameters include the second-order susceptibility $\chi^{(2)}$, the crystal length $L$, and the longitudinal wave vector mismatch $\Delta k_z$.  Minimizing the phase mismatch via angle tuning \cite{karan2020phase} or quasi-phase-matching \cite{torres2004quasi} is crucial to directly increase the down-conversion efficiency and photon pair generation rate.

An important experimental insight is that OAM is conserved between the pump photons and the generated pairs in quasi-collinear configurations \cite{walborn2004entanglement,kopf2025conservation}. Consequently, the OAM of the incoming pump beam is a vital tool to steer the OAM distribution, i.e., the spatial DOF of the signal and idler photons. 

Therefore, we describe both the spatial profile of the photon pair and pump in terms of the Laguerre-Gaussian (LG) modes $\mathrm{LG}^{\ell}_{\pump}(\bm{q}, w)$. As eigenstates of the OAM operator, LG modes provide an infinite yet discrete and measurable basis that allows for the straightforward encoding, measurement, and manipulation of high-dimensional quantum states \cite{fickler2012quantum}. Here, $w$ is the beam waist, $p$ is a positive integer called the radial index. The spiral index $\ell \in \mathbb{Z}$ accounts for the projection of $\ell \hbar$ OAM onto the $z$-axis due to an azimuthally twisted phase distribution \cite{yang2022generation}.

We model the transverse pump spatial distribution by  $V_{\pump}(\bm{q_{\pump}}) = \mathrm{LG}^{\ell_{\pump}}_{p_{\pump}}(\bm{q_{\pump}},w_{\pump})$ and decompose the continuous biphoton state into the discrete LG basis to characterize the OAM distribution of the signal and idler photons:
\begin{equation}
    \ket{\psi} = \sum_{p_{\signal},\ell_{\signal}}  \sum_{p_{\idler},\ell_{\idler}} C_{p_{\signal},p_{\idler}}^{\ell_{\signal},\ell_{\idler}} \biggl( \ket{p_{\signal},\ell_{\signal},w_{\signal}} \otimes \ket{p_{\idler},\ell_{\idler},w_{\idler}} \biggl),
    \label{LGdecomposition}
\end{equation}
where $\ket{p,\ell,w} = \int d\bm{q} \;  \mathrm{LG}_{p}^{\ell}(\bm{q},w) \; \hat{a}^{\dagger}(\bm{q}) \ket{\mathrm{vac}}$ is the single-photon state representations of an LG modes. The parameters $w_{\signal}$ and $w_{\idler}$ are the modes sizes of signal and idler. The expansion amplitude is obtained by
\begin{align}
    &C^{\ell_{\signal},\ell_{\idler}}_{p_{\signal},p_{\idler}} \nonumber = \Bigl( \bra{p_{\signal},\ell_{\signal}} \otimes \bra{p_{\idler},\ell_{\idler}} \Bigl) \ket{\psi} \nonumber \\
    & = \; \mathcal{N}  \iint d\bm{q}_{\signal} \; d\bm{q}_{\idler} \; \int_{-\frac{L}{2}}^{\frac{L}{2}} dz  \;  \chi^{(2)}(z)  \; \mathrm{e}^{ i \Delta k_z(\bm{q}_{\signal},\bm{q}_{\idler}) z} \nonumber \\ 
    & \small{\times} \; \mathrm{LG}_{p_{\mathrm{p}}}^{\ell_{\pump}}(\bm{q}_{\signal}+\bm{q}_{\idler}, w_{\pump}) \Bigl[ \mathrm{LG}_{p_{\signal}}^{\ell_{\signal}}(\bm{q}_{\signal},w_{\signal})\Bigl]^* \Bigl[ \mathrm{LG}_{p_{\idler}}^{\ell_{\idler}} (\bm{q}_{\idler},w_{\idler}) \Bigl]^*,
\label{OverlapAmplitudes}
\end{align}
and derivation details of these amplitude are given in Appendix \ref{Appendix: Fidelity bounds reason}.
 
The projection probability to find the signal photon with $\ell_{\signal}\hbar$ and idler with $\ell_{\idler}\hbar$ OAM is given by
\begin{equation}
    P^{\ell_{\signal},\ell_{\signal}} = \sum_{p_{\signal},p_{\idler}} |C^{\ell_{\signal},\ell_{\idler}}_{p_{\signal},p_{\idler}}|^2.
    \label{PostselectionFreeProbability}
\end{equation}
The probability distribution \eqref{PostselectionFreeProbability} in terms of $\ell_{\signal}$ and $\ell_{\idler}$ is also called the OAM spectrum. 

\begin{figure}[t]
    \centering
    \includegraphics[width=\linewidth]{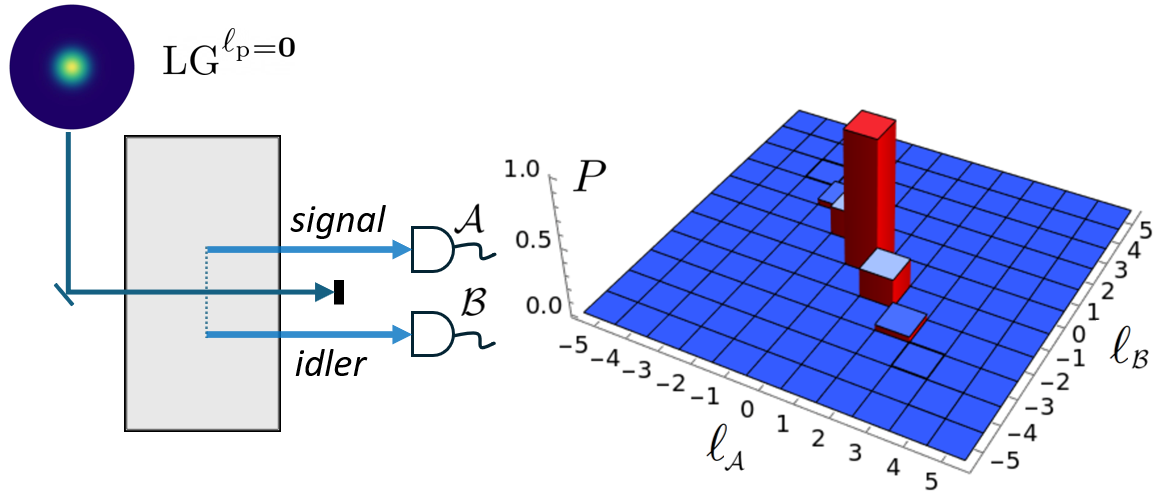}
    \caption{Sketch of a SPDC setup and the corresponding normalized OAM spectrum. A pump beam carrying OAM $\ell_{\pump}=0$ triggers the generation of a photon pair along paths $\mathcal{A}$ and $\mathcal{B}$. The pump is blocked afterwards. The generated OAM distribution of signal and idler photons represents a high-dimensional entangled quantum state. Since the OAM is conserved, the OAM spectrum shows an anti-diagonal line of possible modes whose OAMs add up to zero. The projection probability is in general non-uniformly distributed, as can be seen from the bars of different heights.}
    \label{fig01}
\end{figure}

\subsection{Projection and Fidelity of OAM States}

In most applications, high-dimensional states are measured projectively \cite{mafu2013higher,sevilla2024spectral}, often using holograms on spatial light modulators (SLMs). These devices convert an OAM mode into a Gaussian mode, enabling efficient coupling into a single-mode fiber (SMF) \cite{forbes2016creation}. Nevertheless, the precise experimental measurement of the radial part of LG modes ($p>0$) is challenging due to their complex transverse structures \cite{salakhutdinov2012full,valencia2021entangled}, which can drastically increase the number of required measurements \cite{agnew2011tomography}. Therefore, in many experiments, it is common to project only LG modes with $p = 0$, even though this choice discards a portion of the photon pairs \cite{karan2023postselection}. The projection results in considering only the first term in Eq. \eqref{PostselectionFreeProbability}, so we redefine
\begin{equation}
\ket{\psi^{\prime}} = \sum_{\ell_{\signal},\ell_{\idler}} C^{\ell_{\signal},\ell_{\idler}}_{0,0} \ket{\ell_{\signal}} \otimes \ket{\ell_{\idler}} :=\sum_{\ell_{\signal},\ell_{\idler}} C^{\ell_{\signal},\ell_{\idler}} \ket{\ell_{\signal}} \otimes \ket{\ell_{\idler}}
\label{StateProjection}    
\end{equation}
and only concentrate on the entanglement of the spiral index of the photons.

Due to OAM conservation, the state can be written as a single sum 
\begin{equation}
\ket{\psi^{\prime}} = \sum_{\ell=-\infty}^{\infty} C^{\ell,\ell_{\pump}-\ell} \; \ket{\ell}_{\mathcal{A}} \otimes \ket{\ell_{\pump}-\ell}_{\mathcal{B}}.
\label{OAMconservedState}
\end{equation}
Here, the subscripts $\mathcal{A}$ and $\mathcal{B}$ denote the propagation paths of the signal and idler photons, respectively, as illustrated in Fig. \ref{fig01}. We adopt this notation throughout the paper for all experimental setups. The figure shows a typical OAM spectrum generated by a Gaussian pump beam $\ell_{\pump}=0$. As a consequence of OAM conservation, the down-converted modes are created along an anti-diagonal $\ell_{\mathcal{A}}= \ell_{\pump} -\ell_{\mathcal{B}}$. Since these modes have different amplitudes, they do not form a MES. An MES, in contrast, is characterized by target modes with uniform projection probabilities (bars of equal height) and zero probability for all other modes \cite{bernecker24high}.

To quantify the overlap between the generated OAM state $\ket{\psi^{\prime}}$ and a maximally entangled target  state $\ket{\Psi_{\mathrm{tar}}}$, we employ the quantum fidelity $\mathcal{F}$, defined as $\mathcal{F} = |\braket{ \Psi_{\mathrm{tar}} | \psi^{\prime}}|^2$. This metric provides a measure of the \textit{closeness} between the two quantum states, spanning a range from  $\mathcal{F}=0$ (for orthogonal states) to $\mathcal{F}=1$ (for identical states up to a global phase factor). Therefore, an ideal OAM spectrum for an MES is characterized by a fidelity of $\mathcal{F}=1$, which indicates that all OAM contributions are exclusively concentrated on the target modes. For our further analysis, we need to distinguish between the fidelity of the full Hilbert space $\mathcal{F}$ and the subspace fidelity $\mathcal{F}_{\mathrm{sub}}$. The latter is calculated when the summation in Eq. \eqref{OAMconservedState} is restricted to a specific range of OAM values $\ell$.

\section{Engineering of the OAM spectrum}

In our work, we aim to shape the OAM spectrum to generate high-fidelity, maximally entangled target states. This section discusses relevant quantum state engineering methods used to alter the OAM probability distribution of photon pairs.

\subsection{Spatially engineered pump beams}
\label{Spatially engineered pump beams}

SLMs have revolutionized beam shaping, enabling both the measurement and creation of exotic light fields \cite{bolduc2013exact}. This advancement has opened new perspectives for tailoring the pump beam in SPDC. A suitable hologram can be programmed onto an SLM to generate an arbitrary superposition of LG pump modes with various OAM values:
\begin{equation}
    V_{\pump}(\bm{q}_{\pump}) = \sum_{\ell_{\pump}} a_{\ell_{\pump}}  \; \mathrm{LG}_{0}^{\ell_{\pump}}(\bm{q}_{\pump},w_{\pump}).
    \label{PumpinLGsuperposition}
\end{equation}
We again set all radial indices $p_{\pump}=0$, since the OAM spectrum of signal and idler photons can be shaped by directly adjusting the spiral index of the LG pump mode. The holograms provide independent control over each weight $a_{\ell_{\pump}}$, allowing us to program any desired superposition of LG pump modes. Consequently, when the pump beam \eqref{PumpinLGsuperposition} is used in SPDC, it produces a photon pair in the following state:
\begin{equation*}
\ket{\psi} = \sum_{\ell=-\infty}^{\infty} \sum_{\ell_{\pump}} a_{\ell_{\pump}}C^{\ell,\ell_{\pump}-\ell} \; \ket{\ell}_{\mathcal{A}} \otimes \ket{\ell_{\pump}-\ell}_{\mathcal{B}}.
\end{equation*}
Next to the pump beam waist $w_{\pump}$, the pump weights $a_{\ell_{\pump}}$\textbf{ }provide a powerful tool for manipulating the OAM distribution of the entangled photon pair. The resulting OAM spectra are straightforward to visualize. Each pump mode contributes to OAM conservation, resulting in multiple anti-diagonals in the OAM spectrum of the generated photon pairs.

\begin{figure}[t]
    \centering
    \includegraphics[width=0.95\linewidth]{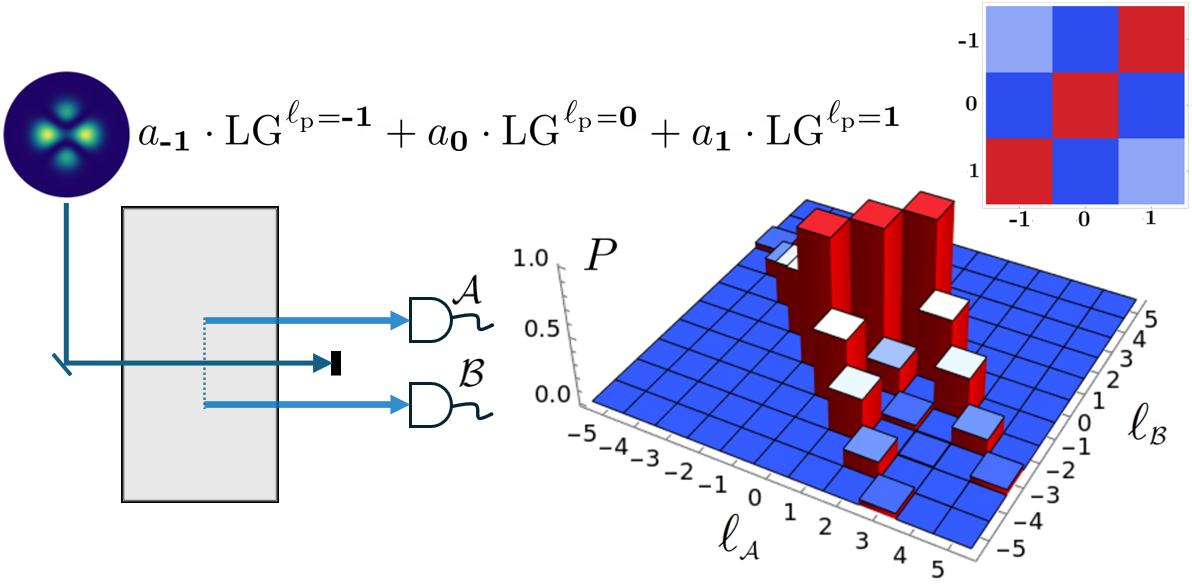}
    \caption{Sketch of a SPDC setup and the corresponding normalized OAM spectra. Suppose a superposition of different OAM modes pumps the nonlinear crystal. In that case, the OAM spectrum of the photon pair spans multiple anti-diagonals, ensuring that the conservation law is fulfilled for each term in the superposition. The OAM spectrum can be adjusted by the additional parameters $a_{\ell_{p}}$. Here, for example, a superposition of OAM indices $\ell_{\pump} = -2, 0, 2$ is chosen such that the biphoton OAM modes $\ket{-1,-1}, \ket{0,0}$ and $\ket{1,1}$ have the same detection probability. The fidelity for the state $\ket{\psi_{\mathrm{tar}}} =1/\sqrt{3} \Bigl( \ket{-1,-1}+ \ket{0,0}+\ket{1,1} \Bigl)$ in the subspace $\ell=\{-1,0,1\}$ (shown in the inset) is $\mathcal{F}_{\mathrm{sub}}=0.876$. However, a large amount of modes are concentrated outside this subspace, which is quantified by the fidelity $\mathcal{F}=0.401$ for the whole space. }
    \label{fig02}
\end{figure}

Fig.  \ref{fig02} illustrates this concept for a pump beam superposition of $\ell_{\pump} = -2, 0, 2$. By appropriately choosing the pump coefficients $a_{\ell_{\pump}}$, the OAM spectrum can be engineered to populate multiple anti-diagonals simultaneously and to equalize the distribution of the down-converted OAM modes. For instance, setting $a_{-2}/a_0$ and $a_{+2}/a_0$ appropriately equalizes the amplitudes for the modes  $ \{\ket{-1,-1}, \ket{0,0}, \ket{1,1} \}$. These weights can be found either analytically \cite{baghdasaryan2022generalized, bernecker24high} or using stochastic algorithms \cite{kovlakov2018quantum}.

The fidelity of the MES $\ket{\Psi_{\mathrm{tar}}}=\frac{1}{\sqrt{3}} \Bigl( \ket{-1,-1} + \ket{0,0} + \ket{1,1}  \Bigl)$  is $\mathcal{F}_{\mathrm{sub}}=0.876$ when considering only modes within the $\ell_{\signal}=\ell_{\idler}=-1,0,1$ subspace (as depicted in the inset in Fig. \ref{fig02}), while all other modes are disregarded. In contrast, the global fidelity is significantly lower at  $\mathcal{F}=0.401$, a consequence of the large OAM distribution.

\subsection{OAM and Phase Shifts }

\begin{figure}[t]
    \centering
    \includegraphics[width=\linewidth]{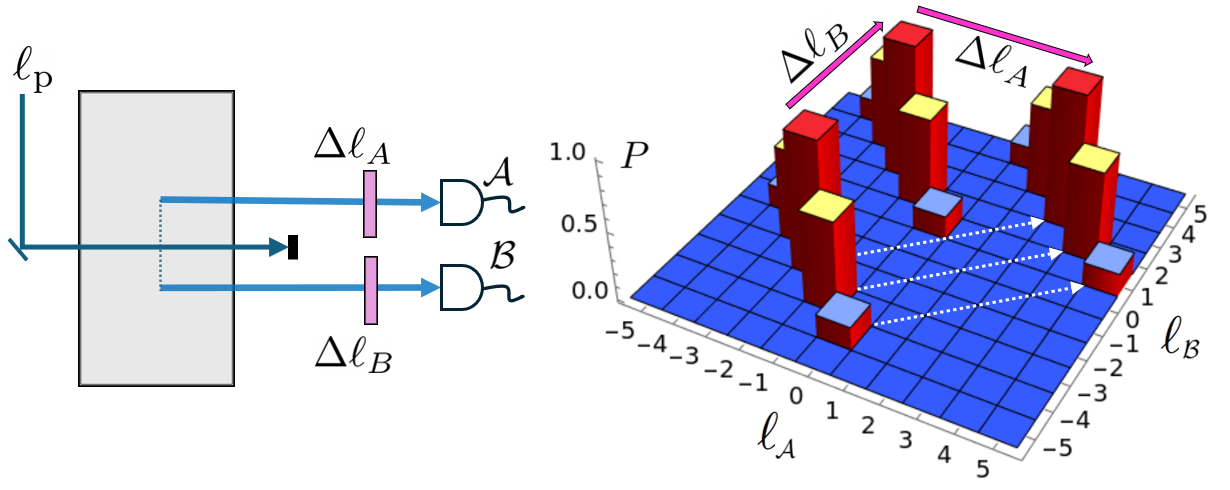}
    \caption{Mode transformation via OAM shifts. After generation, the OAM of photons traveling along paths $\mathcal{A}$ and $\mathcal{B}$ can be manipulated by adding or subtracting $\Delta\ell_{A}$ or $\Delta\ell_{B}$ quanta of OAM, respectively. This is achieved using components such as spiral phase plates. In collinear configurations, the transformation is typically $\Delta\ell_{A}=\Delta\ell_{B}$ as both photons pass through the same device.  Non-identical shifts $\Delta\ell_{A} \neq \Delta\ell_{B}$ are possible when the components are placed after the photon pair has been separated. The action of applying $\Delta\ell_{A}=4$ and $\Delta\ell_{B}=5$ is illustrated on the right for an OAM spectrum generated by an initial pump mode $\ell_{\pump}=-4$. These mode shifts affect all OAM modes simultaneously, displacing the entire anti-diagonal.}
    \label{fig03}
\end{figure}

\begin{figure*}
    \centering
    \includegraphics[width=0.85\linewidth]{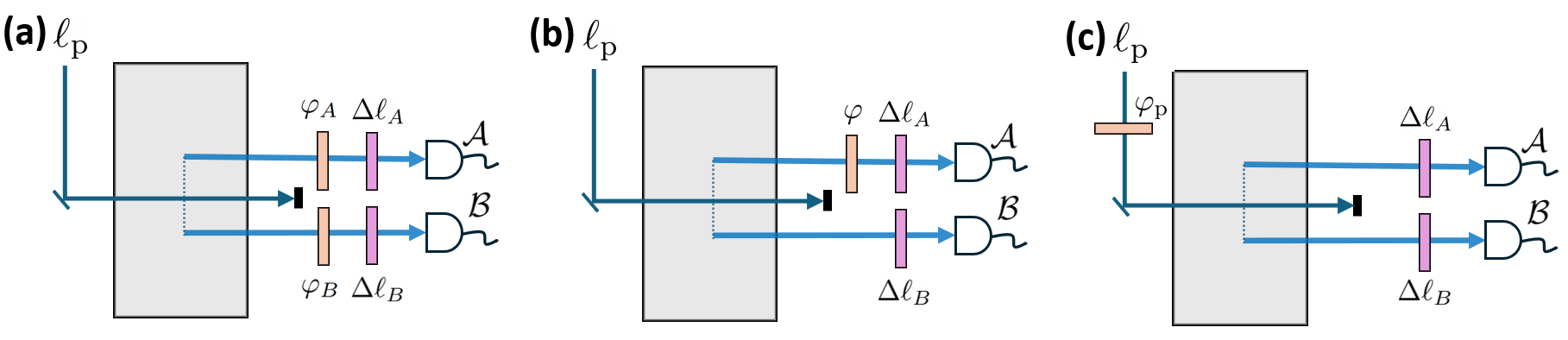}
    \caption{Phase transformations via phase shifters. Very similar to the mode shift concept in Fig. \ref{fig03}, the optical phase can be altered along paths $\mathcal{A}$ and $\mathcal{B}$. (a) Shifters add a phase factor $\varphi_A$ along path $\mathcal{A}$ and $\varphi_B$ along $\mathcal{B}$, which effectively adds a global phase factor of $\varphi := \varphi_A + \varphi_B$  for the two-photon state. (b) Such a shift is mathematically equally obtained by a shift $\varphi$ only in one of the paths, or (c) inherited from the pump beam, if the pump beam undergoes a phase shift as well, e.g., via an SLM.}
    \label{fig04}
\end{figure*}

There are several ways to change the OAM of generated photons. A spiral phase plate (SPP) \cite{beijersbergen1994helical} is a cylindrical component with an azimuthally varying thickness that imprints a phase retardation, twisting the wavefront. The OAM value changes by $\Delta \ell$  depending on the SPPs step height and the operating wavelength. The device can either add or subtract OAM depending on the handedness of its design. In addition to SLMs and SPPs, other components manipulate the OAM of photons include generalized SPPs \cite{khonina2020properties} and q-plates, which convert spin angular momentum into OAM \cite{rubano2019q}.

The action of an SPP operation can be written as $\bm{\Delta} \bm{\ell} \ket{\ell} = \ket{\ell + \Delta \ell}$, where the operator $\bm{\Delta} \bm{\ell}$ acts on the OAM mode representation $\ket{\ell}$ as 
\begin{align}
    \ket{\ell} \mapsto & \int d\bm{q} \; \underbrace{\mathrm{LG}_{p=0}^{\ell}(\bm{q},w) \; \mathrm{e}^{i \Delta \ell \mathrm{Arg}(\bm{q})}}_{\mathrm{LG}_{p=0}^{\ell+\Delta \ell}(\bm{q},w)} \; \hat{a}^{\dagger}(\bm{q}) \ket{\mathrm{vac}} \nonumber \\
    &= \ket{\ell+\Delta \ell}
    \label{OAMshiftexpression}
\end{align}
This transformation is valid only for the lowest radial order $p=0$, as assumed in our analysis through projection onto $p_{\signal} = p_{\idler} = 0$. If all radial modes are considered (see Eq. \eqref{PostselectionFreeProbability}), the SPP may generate a mixture of LG modes with the intended OAM index but different radial numbers \cite{hakimi2025detection}.

Since the biphoton state is in superposition of different OAM modes, mode shifting elements which add or subtract $\Delta \ell_A$ OAM in path $\mathcal{A}$ and $ \Delta \ell_B$ OAM in path $\mathcal{B}$, would affect the state as
\begin{equation*}
\ket{\psi^{\prime}} = \sum_{\ell=-\infty}^{\infty} C^{\ell,\ell_{\pump}-\ell} \; \ket{\ell + \Delta \ell_A }_{\mathcal{A}} \otimes \ket{\ell_{\pump}-\ell + \Delta \ell_B }_{\mathcal{B}}. 
\end{equation*}

Fig.  \ref{fig03} illustrates the same setup as before, but with mode shifters inserted along paths $\mathcal{A}$ and $\mathcal{B}$. One possible implementation employs two different SPPs, each placed in a separate arm after the photon pair is deterministically split by a polarizing beam splitter or dichroic mirror. Alternatively, in a collinear configuration, a single mode-shifting element can be used to apply the same phase shift to both photons $\Delta \ell_A = \Delta\ell_B$. 

Such an operation can, in principle, shift any mode on the anti-diagonals to a desired spot in the OAM spectrum as exemplified in Fig. \ref{fig03}. While the operation shifts all modes simultaneously, the height of the mode probability remains unchanged. 

The optical phase of the down-converted photons can also be adjusted using phase-shifting components such as wave plates (QHQ scheme \cite{kysela2020path}), tilting glass wedges \cite{thomas2010measurement}, piezo-actuated mirrors \cite{briles2010simple}, or trombone systems \cite{lualdi2025fast}. Fig.  \ref{fig04} (a) shows a phase shifter inserted into the previous setup. Each path introduces a phase delay, $\varphi_A$ and $\varphi_B$, respectively, which enters the quantum state as exponential multiplicative factors:
\begin{small}
\begin{align}
\ket{\psi^{\prime}} &= \sum_{\ell=-\infty}^{\infty} C^{\ell,\ell_{\pump}-\ell} \; \mathrm{e}^{i \varphi_A} \ket{\ell +  \Delta \ell_A }_{\mathcal{A}} \otimes  \mathrm{e}^{i \varphi_B} \ket{\ell_{\pump}-\ell+  \Delta \ell_B}_{\mathcal{B}} \nonumber \\ 
&= \sum_{\ell=-\infty}^{\infty} C^{\ell,\ell_{\pump}-\ell} \; \underbrace{\mathrm{e}^{i\varphi_A+ \varphi_B}}_{:=\mathrm{e}^{i \varphi}} \ket{\ell +  \Delta \ell_A }_{\mathcal{A}} \otimes \ket{\ell_{\pump}-\ell+ \Delta \ell_B}_{\mathcal{B}} \nonumber,
\end{align}
\end{small}
which corresponds to a total phase shift $\varphi$ only in one path, see in Fig.  \ref{fig04} (b). A phase shift applied to the pump beam before down-conversion can also introduce a phase for the biphoton state, as illustrated in Fig.  \ref{fig04} (c) using a SLM \cite{liu2018coherent}. The phase shift is not relevant in a single-crystal configuration, where it vanishes upon projection, but is crucial for introducing a relative phase between different crystals in path-identity setups.

\subsection{Path Identity}

The principle of path identity in multi-crystal setups relies on the indistinguishability of the photon-pair generation process \cite{wang1991induced, herzog1994frustrated, krenn2017entanglement}. When a laser pumps multiple identical crystals aligned to emit into the same spatial modes, the resulting quantum state is a coherent superposition of all possible generation events, assuming only one photon pair is created. In other words, quantum interference is achieved by matching all DOFs of the generated photons, thereby eliminating any \textit{which-crystal} information that could distinguish their origin.

\begin{figure}[b]
    \centering
    \includegraphics[width=0.9\linewidth]{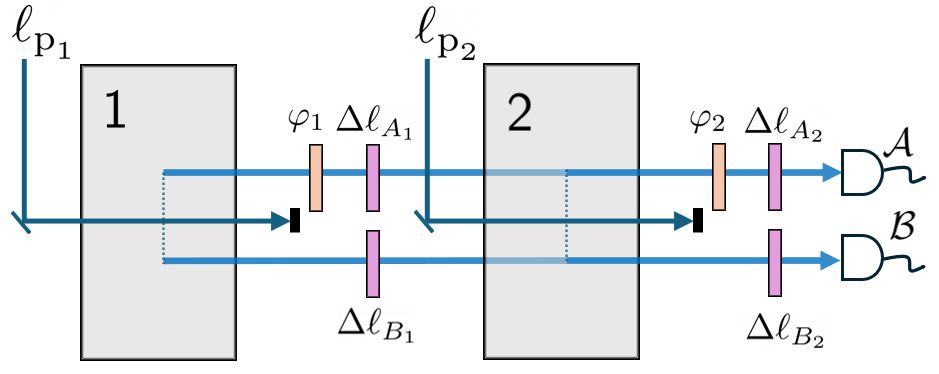}
    \caption{Two-crystal path identity setup. The paths from crystal $\mathsf{1}$ are made identical (perfectly overlapped) with the paths $\mathcal{A}$ and $\mathcal{B}$ from crystal $\mathsf{2}$. Both crystals are coherently pumped with OAM values $\ell_{p_1}$ and $\ell_{p_2}$. The setup can be built by using a single pump beam consecutively (i.e., without blocking the pump). Alternatively, it can be realized interferometrically, where the initial pump beam is split and its OAM is modified differently on each path before being directed to the crystals. Any distinguishability between the photon pairs generated in crystals $\mathsf{1}$ and $\mathsf{2}$, such as misalignment of the paths, will reduce the quantum interference of different generation events. By incorporating the transformation components discussed earlier for the OAM and phase, we can directly engineer high-dimensional quantum states.}
    \label{fig05}
\end{figure}

A simple realization of path identity can be realized in a two-crystal configuration. Consider merging two single-crystal setups and overlapping the paths from crystal $\mathsf{2}$ with those from crystal $\mathsf{1}$, as sketched in Fig. \ref{fig05}, placing the pair in a superposition of origins. Each crystal is pumped by a distinct OAM mode: crystal $\mathsf{1}$ by $\ell_{p_1}$ and crystal $\mathsf{2}$ by $\ell_{p_2}$. 

Next to their paths, also the joint spectral amplitude \cite{shukhin2024two}, and polarization of the crystals and must match. For this reason, these setups typically use identical nonlinear crystals, which ensures consistent phase-matching conditions for polarization and a shared spectral range for the down-converted photons. To ensure temporal indistinguishability in the setup, two conditions must be met: first, the arrival times of the signal and idler photons must be indistinguishable, and second, no temporal information can reveal whether the pair originated in crystal $\mathsf{1}$ or crystal $\mathsf{2}$. Appendix \ref{tempindist} provides a detailed account for these conditions.

We explicitly operate in the low-gain regime, ensuring that only a single photon pair is present at any given time in the setup. At higher gain, simultaneous multi-pair emissions may occur \cite{takeoka2015full}. In the absence of photon-number-resolving detection, these higher-order contributions act as noise and reduce the coherence of the photon-pair generation processes.

In Fig. \ref{fig05}, mode shifts  $\Delta \ell_{A_j,B_j}$ and phase shifts $\varphi_j$ are also included after each crystal, $j \in \{ 1,2\}$. To independently manipulate the photons along paths $\mathcal{A}$ and $\mathcal{B}$ with optical components, these paths may first be spatially separated and then recombined to exactly overlap with the corresponding paths from crystal $\mathsf{2}$.

For the purposes of this paper, we will assume perfectly overlapped paths and lossless optical components. Under our assumptions, the resulting quantum state after the second crystal is a coherent superposition of the states produced in each crystal:
\begin{equation}
    \ket{\Psi} = \frac{1}{\sqrt{N}}  \Bigl( \ket{\Psi_{\mathsf{1}}} +  \mathcal{X} \cdot   \ket{\Psi_{\mathsf{2}}} \Bigl),
    \label{PIEtwocrystals}
\end{equation}
where each individual state is given by
\begin{widetext}
\begin{align*}
\ket{\Psi_{\mathsf{1}}} &= \sum_{\ell_1=-\infty}^{\infty} C^{\ell_1,\ell_{p_1}-\ell_1} \; \mathrm{e}^{i ( \varphi_1 + \varphi_2 )} \; \ket{\ell_1 + \Delta \ell_{A_1}+  \Delta \ell_{A_2}}_{\mathcal{A}} \otimes \ket{\ell_{p_1}-\ell_1+  \Delta \ell_{B_1}+  \Delta \ell_{B_2}}_{\mathcal{B}} \nonumber \\
\ket{\Psi_2} &= \sum_{\ell_2=-\infty}^{\infty} C^{\ell_2,\ell_{p_2}-\ell_2} \; \mathrm{e}^{i \varphi_2} \; \ket{\ell_2 +   \Delta \ell_{A_2}}_{\mathcal{A}} \otimes \ket{\ell_{p_2}-\ell_2+  \Delta\ell_{B_2}}_{\mathcal{B}}.
\end{align*}
\end{widetext}

The normalization is ensured by the constant
\begin{align*}
    N &=  \sum_{\ell_1} |C^{\ell_1,\ell_{p_1}-\ell_1}|^2 +  \mathcal{X}^2 \cdot  \sum_{\ell_2} |C^{\ell_2,\ell_{p_2}-\ell_2}|^2.
\end{align*}
In the provided setup, the photon pair originating from crystal $\mathsf{1}$ undergoes two sequential OAM and phase shifts. The pair from crystal $\mathsf{2}$ is subject to only one state transformation. The factor $\mathcal{X}$ is used to tune the relative pump strengths between crystal $\mathsf{1}$ and $\mathsf{2}$ as $\mathcal{P}_2=\mathcal{X} \cdot \mathcal{P}_1$ with equal strengths when $\mathcal{X} = 1$. 

The total state can be controlled in several ways: by shaping the OAM pump mode of individual crystals, by altering the pump strength ratio with $\mathcal{X}$, and by shifting the OAM mode distribution in between. In the ideal case, these operations allow for an increase in the dimension of the state. A generalization to $n$-crystal setups can be found in Appendix \ref{n crystal scheme}.

As an example, we consider a scenario where two crystals are pumped with equal strength. One crystal produces the OAM state $\ket{0,0}$, while the other produces the state $\frac{1}{\sqrt{2}} (\ket{0,1} + \ket{1,0})$. Through path identity, the total state is a coherent superposition of the outputs from the two crystal, resulting in a normalized state given by:
\begin{equation*}
    \ket{\Psi} = \frac{1}{\sqrt{2}} \ket{0,0 } + \frac{1}{2} \Bigl(\ket{0,1} + \ket{1,0} \Bigl),
\end{equation*}
a partially separable state with non-equal amplitudes. 

The balance of the superposition can be adjusted by changing the relative pumping strength to $\mathcal{X} = \sqrt{2}$; the pump power for crystal $\mathsf{2}$ is increased relative to crystal $\mathsf{1}$. This can be achieved experimentally using an unbalanced beam splitter to pump in average crystal $\mathsf{2}$ more. If the photons from crystal $\mathsf{1}$ experience an addition of OAM $\Delta \ell_{A_1}= \Delta \ell_{B_1}=2$, we prepare the three-dimensional MES
\begin{equation*}
    \ket{\psi} = \frac{1}{\sqrt{3}} \Bigl( \ket{2,2 } + \ket{0,1} + \ket{1,0} \Bigl).
\end{equation*}

\begin{figure*}
    \centering
    \includegraphics[width=\linewidth]{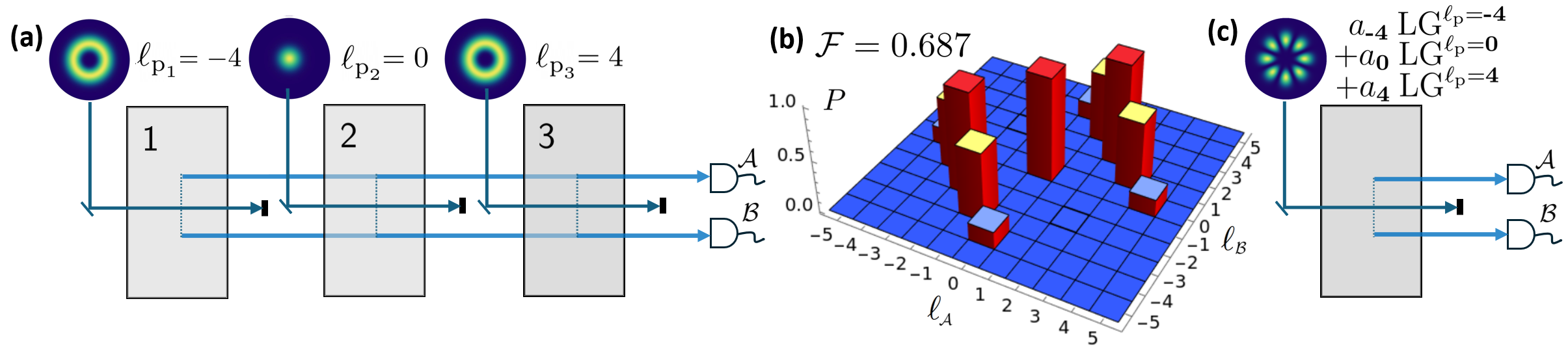}
    \caption{Two different methods to generate high-dimensional entangled states with the same OAM distribution. (a) Entanglement by path identity using three crystals $\mathsf{1,2,3}$, each coherently pumped by a laser beam that was split into three parts beforehand. In each crystal, a different pump OAM mode ($\ell_{\pump} = -4, 0, 4$, respectively) triggers the generation of a photon pair. (b) The detection setup is configured such that the crystal pumped with $\ell_{\pump} = 0$ produces only the biphoton OAM mode $\ket{0,0}$. The other two crystals generate the modes $\ket{-2,-2}$ and $\ket{2,2}$, as well as a broader range of OAM modes satisfying the OAM conservation. As a result, the fidelity of the MES $\ket{\psi_{\mathrm{tar}}} = 1/\sqrt{3} \Bigl( \ket{-2,-2} + \ket{0,0} + \ket{2,2} \Bigr)$ over the full space is $\mathcal{F} = 0.687$. By post-selecting the subspace $\ell = {-2, 0, 2}$, the fidelity is $\mathcal{F}_{\mathrm{sub}} = 1$. (c) The same OAM spectrum shown is prepared using an engineered pump superposition with suitably weighted OAM modes $\ell_{\pump} = -4, 0, 4$.}
    \label{fig06}
\end{figure*}

\section{Results and Discussion} 

This chapter establishes the equivalence between the state in Eq. \eqref{PIEtwocrystals} and the biphoton states generated by the spatially engineered pumps from Chapter \ref{Spatially engineered pump beams}. We will also identify the conditions under which the fidelity for MESs is less than one and discuss the resulting limitations for generating genuine MESs. Unless otherwise specified, our analysis assumes identical $L=10$ mm type-II ppKTP crystals, pumped by a 405 nm continuous-wave laser. The configuration produces degenerate $k_{\signal}=k_{\idler}$ photon pairs, spectrally centered around 810 nm. The parameters used are adapted from Ref. \cite{kysela2020path}.

\subsection{Equivalence to Pump Engineering}
\label{equivalence}

In path-identity approach, the relative pump power between crystals serves as a new tuning parameter to engineer OAM spectra. Additionally, using different OAM pump modes $\ell_{p_1}, \ell_{p_2}, ... $ or applying specific OAM shifts can cause the final state to occupy multiple anti-diagonals in the joint OAM spectrum. Adjusting the relative pump power between crystals pumped by different OAM modes is analogous to adjusting the weights of a pump superposition in Eq. \eqref{PumpinLGsuperposition}.

In the following example, we show that path identity approach and the use of spatially shaped pump Eq. \eqref{PumpinLGsuperposition} superpositions are equivalent methods in terms of the resulting OAM spectrum. Consider the generation of the maximally entangled target state 
\begin{eqnarray*}
    \ket{\psi_{\mathrm{tar}}} = \frac{1}{\sqrt{3}} \Bigl( \ket{-2,-2}+\ket{0,0} + \ket{2,2}\Bigl).
\end{eqnarray*}

The most straightforward path identity setup to generate these three-dimensional MES includes three crystals, $\mathsf{1, 2 \; \& \;3}$, pumped by OAM modes $\ell_{p_1}=-4, \ell_{p_2}=0$, and $\ell_{p_3}=4$, respectively. This scenario is sketched in Fig.  \ref{fig06} (a). The relative pump strength for crystal $\mathsf{1}$ and $\mathsf{3}$ can be set to $\mathcal{X}\approx 1.72$, where  $\mathcal{P}_2= \mathcal{X} P_1= \mathcal{X} P_3$. This sets the  photon pair creation of the biphoton modes $\ket{-2,-2}, \ket{0,0}, \ket{2,2}$ equally likely. Nevertheless, OAM conservation also lead to the population of several \textit{unintended} modes, those not contributing to the desired target state. In Fig. \ref{fig06} (b), a realistic OAM spectrum is shown for $w_{\pump}=15 \mu$m and $w_{\signal}=w_{\idler}=25 \mu$m. 

Here, for the diagonal $\ell_{p_2}=0$, most of the amplitude is concentrated on $\ket{0,0}$, but for the diagonals $\ell_{p_1}=-4$ and $\ell_{p_3}=4$, there are non-vanishing contributions from $\ket{-1,-3}, \ket{-3,-1}$ and $\ket{1,3}, \ket{3,1}$. In the next section, we will see that the amplitude ratios of these modes to $\ket{2,2}$ and $\ket{-2,-2}$ are fixed. In turn, this leads to a reduced fidelity over the whole OAM space, $\mathcal{F}=0.687$. When the photons are projected only in the subspace $\ell \in \{ -2,0,2 \}$, we find $\mathcal{F}_{\mathrm{sub}}=1$, but leave a significant amount of pairs unused.

The same OAM spectra can also be generated by a single crystal pumped with an LG superposition of
\begin{equation}
    V_{\pump}= a_{-4} \cdot\mathrm{LG}_{0}^{-4} + a_{0} \cdot\mathrm{LG}_{0}^{0} + a_{4} \cdot\mathrm{LG}_{0}^{4},
\end{equation}
by setting $a_{-4}/a_{0} = a_{+4}/a_{0}$ as depicted in Fig. \ref{fig06}  (c). Since the OAM spectra for (a) and (c) are the same, the fidelity is equally $\mathcal{F} = 0.687$. The experimental effort for entanglement by path identity requires coherently setting up the three crystals, while ensuring that at most one photon pair is generated simultaneously. Using the spatially engineered pump in only one crystal avoids these problems. 

\begin{figure}[b]
    \centering
    \includegraphics[width=\linewidth]{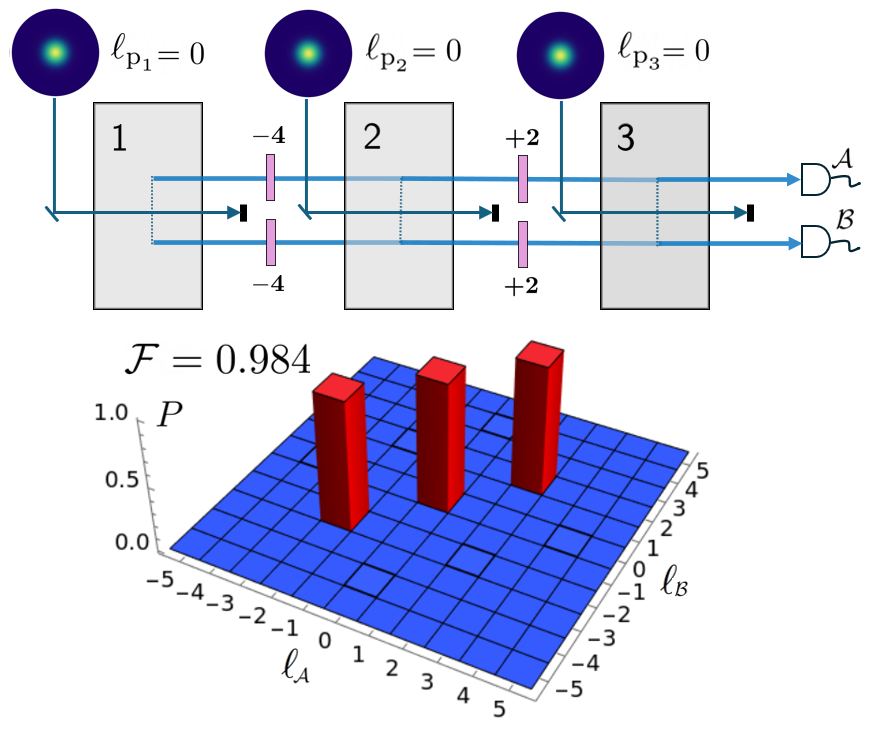}
    \caption{An alternative path identity setup to generate the target MES $ \ket{\psi_{\mathrm{tar}}} = \frac{1}{\sqrt{3}} \Bigl( \ket{-2,-2}+\ket{0,0} + \ket{2,2}\Bigl)$. In comparison to Fig.  \ref{fig06}, three coherently pumped nonlinear crystals by a Gaussian mode $\ell_{\pump}$ produces the state $\ket{0,0}$ and mode shifters are used to create a superposition of distinct OAM modes. Since this approach does not generate unintended side modes due to OAM conservation, since we are capable to concentrate most of the amplitude solely on $\ket{0,0}$, the fidelity is much higher, $\mathcal{F}=0.984$}
    \label{fig07}
\end{figure}

\begin{figure*}
    \centering
    \includegraphics[width=0.9\linewidth]{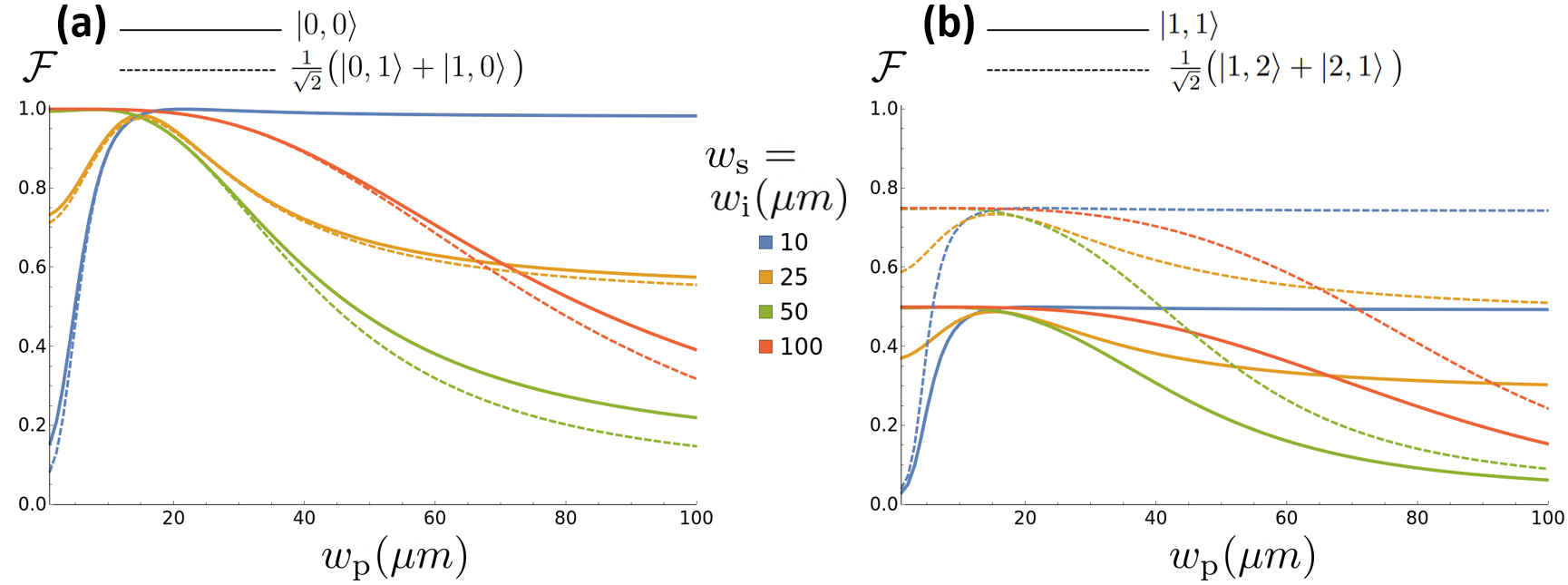}
    \caption{Tuning of the pump $w_{\pump}$ and collection waists $w_{\signal}, w_{\idler}$ for maximal fidelity in a single nonlinear crystal of different exemplary target states. We consider a range of typical waist parameters commonly used in experiments involving OAM modes of light. (a) The target states $\ket{0,0}$ (solid line) and $\frac{1}{\sqrt{2}} (\ket{0,1} + \ket{1,0})$ (dashed line) exhibit similar fidelity behavior and can be optimized to reach maximal fidelity $\mathcal{F} = 1$ through appropriate waist tuning. (b) In contrast, the fidelity for the target states $\ket{1,1}$ (solid line) and $\frac{1}{\sqrt{2}} (\ket{1,2} + \ket{2,1})$ (dashed line) is significantly lower and cannot exceed bounds $\mathcal{F} < 1$. The fidelity for pump modes with $\ell_{\pump} = 2$ ($\ell_{\pump} = 3$) is limited by the generation of additional modes, like $\ket{0,2}, \ket{2,0}$ ($\ket{0,3}, \ket{3,0}$) which are prepared alongside the target state. These \textit{unintended} modes are present independently of the values of the waist sizes and therefore reduce the achievable fidelity. This limitation does not apply to the target states in (a), where all undesired modes can be effectively suppressed by appropriate waist tuning.} 
    \label{fig08}
\end{figure*}

An alternative design uses path identity to pump all three crystals with the same OAM mode, $\ell_{p} = 0$, shown in Fig. \ref{fig07}. By carefully choosing the pump and collection waists, the configuration almost exclusively produces the $\ket{0,0}$ mode, thereby avoiding unintended side modes. Then, by inserting OAM shifts of $\Delta \ell_{A_1} = \Delta \ell_{B_1} = -4$, $\Delta \ell_{A_2} = \Delta \ell_{B_2} = 2$, and $ \Delta \ell_{A_3} = \Delta \ell_{B_3} = 0$, the target MES with a full-space fidelity close to unity can be prepared. This choice was avoided in the proof-of-principle experiment from  Ref. \cite{kysela2020path}. Even a slight non-collinear emission of the photons resulted in an imperfect OAM shifts preventing to implement $\Delta \ell_{A_{1}}= \Delta \ell_{B_{1}}$ and $\Delta \ell_{A_{2}}= \Delta \ell_{B_{2}}$ with \textit{one} SPP, respectively. Overcoming this challenge would require a more sophisticated interferometric setup, using type-0 or type-I SPDC for a perfect collinear emission \cite{lee2025optimizing} or a combination of quarter-wave-, half-wave-, and $ q$-plates.

When the crystals are pumped by OAM modes $\ell_{\pump_1}=\ell_{\pump_2}=...=0$, the dimensionality of HDES generated via path-identity approach scales linearly with the number of crystals. For $n$ crystals, $n$ OAM mode shifters placed after each crystal can be used to add another term to the resulting superposition, yielding a state with dimensionality $d=n$. Increasing the dimensionality to $d+1$ requires to add another crystal with a corresponding mode shifter to the setup. A better compromise between achieving a high fidelity and maintaining a simple setup could be found by generating high-fidelity HDES within a single crystal and coherently superposing the entangled states. As we will show in the next section, we are limited in this regard.

\subsection{Fidelity bounds}

Efficient entanglement generation using the path identity approach requires creating high-fidelity target modes in each crystal. These modes are then coherently superposed, as shown with the $\ket{0,0}$ mode in Fig. \ref{fig07}. Therefore, it is necessary to force the amplitude of unintended modes to vanish. However, the intrinsic conservation of OAM places a constraint on the tuning of the expansion amplitudes given in Eq. \eqref{OverlapAmplitudes}. 

Here, we consider symmetric OAM spectra  $C^{\ell_{\signal},\ell_{\idler}} = C^{\ell_{\idler},\ell_{\signal}}$, which is valid under the conditions of equal momenta $k_{\signal} = k_{\idler}$ and collection waits $w_{\signal} = w_{\idler}$. This symmetry is highly beneficial for preparing HDES, as the modification of a given mode results in direct adjustment of its symmetric mode without any further modification. \cite{baghdasaryan2022generalized, bernecker24high}.

For any given OAM mode of the pump, the ratios between all amplitudes $C^{\ell_{\signal},\ell_{\idler}}$ that fulfill either $\ell_{\signal}, \ell_{\idler} \geq 0$ or $\ell_{\signal}, \ell_{\idler} \leq 0$ are fixed, i.e. these modes can be adjusted independently from each other. The ratio cannot be adjusted by changing the pump waist $w_{\pump}$, the collection waists $w_{\signal}, w_{\idler}$ or the crystal length $L$. An analytical proof for this claim is provided in Appendix \ref{Appendix: Fidelity bounds reason}. This limitation affects every occupied anti-diagonal in the OAM spectrum, meaning the fixed amplitude ratios cannot be adjusted by changing relative pump powers or spatially shaped pump beams. 

A pivotal role is filled by the relative Mode Number (RMN) defined by $N_R = |\ell_{\pump}| - |\ell_{\signal}| - |\ell_{\idler}| \in \{ 0, -2, -4, \dots \}$. The RMN has been studied for decoupling spatial and spectral DOFs \cite{baghdasaryan2022generalized} and for enhancing OAM entanglement via phase-matching modifications \cite{bernecker24high}. The RMN is closely related to the fidelity bounds, as all modes with RMN $N_R = 0$ exhibit this untunable property, meaning their relative amplitude ratios are fixed and cannot be adjusted by changing the pump or collection waists. Conversely, modes with $N_R < 0$ are tunable and can be suppressed by appropriately choosing the pump waist $w_{\pump}$ and the collection waists $w_{\signal}=w_{\idler}$. 

The number of modes with fixed amplitude ratios increases with the magnitude of the pump OAM $|\ell_{\pump}|$. For instance, $\ell_{\pump}=2$ generates three modes, $\ket{0,2},\ket{1,1}, \ket{2,0}$, that satisfy $\ell_{\signal}, \ell_{\idler} \geq 0$.  In contrast, $\ell_{\pump}=5$ already yields six modes with the same property: $\ket{0,5}, ... , \ket{5,0}$. These individual modes cannot be adjusted independently due to the fixed amplitude ratio. The permanent presence of unintended modes, which cannot be equalized with target modes, inevitably reduces the fidelity of the target state. The fixed ratios $P^{0, 2}/P^{1,1}= 1/2$ (see Fig. \ref{fig02}) or $P^{1, 3}/P^{2,2}= 2/3$ and $ P^{0, 4}/P^{2,2}=1/6$ (see Fig. \ref{fig06}) significantly reduce the fidelity of the target state.

\begin{figure}[t]
    \centering
    \includegraphics[width=0.95\linewidth]{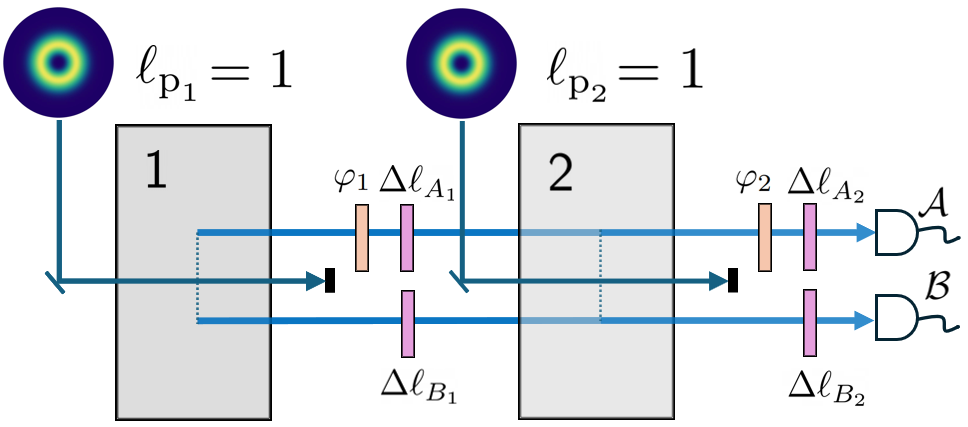}
    \caption{A sketch of a general optimal two-crystal setup. Taking into account the fidelity bounds, the optimal dimensionality for a $d=4$ MES can be achieved in a configuration of two crystals pumped by OAM modes $\ell_{\pump_1}=\ell_{\pump_2}=1$.} 
    \label{fig09}
\end{figure}

The fidelity bounds confine us to a specific set of target states that are achievable with near-unity fidelity in a single crystal. For these states, we can identify target modes satisfying the condition $\ell_{\signal}, \ell_{\idler} \geq 0$ or $\ell_{\signal}, \ell_{\idler} \leq 0$  (which is equivalent to having an RMN of $N_R=0$). All other modes generated by the conservation law $\ell_{\pump}=\ell_{\signal}+\ell_{\idler}$ violate this condition (i.e., have $N_R<0$). By carefully choosing the waist parameters, we can concentrate nearly all the amplitude onto these specific target modes. The target states that meet these criteria are
\begin{align*}
    \ket{\psi_1} &= \ket{0,0} \\
    \ket{\psi_2} &= \frac{1}{\sqrt{2}} \Bigl( \ket{0,1} + \ket{1,0} \Bigl) \\
    \ket{\psi_3} &= \frac{1}{\sqrt{2}} \Bigl( \ket{-1,0} + \ket{0,-1} \Bigl).
\end{align*}
These states can be tuned to achieve fidelities of  $\mathcal{F} \approx 1$. All biphoton states generated by $|\ell_{\pump}|>1$ will have fidelities $\mathcal{F}<1$. Fig.  \ref{fig08} (a) shows the fidelity for the target states $\ket{\psi_1}$ ($\ell_{\pump}=0$) and $\ket{\psi_2}$ ($\ell_{\pump}=1$) as a function of pump and collection waists in an experimentally typical range of waist sizes. Due to the property $C^{\ell_{\signal},\ell_{\idler}}=(C^{-\ell_{\signal},-\ell_{\idler}})^*$ ( (see Appendix \ref{Appendix: Fidelity bounds reason} or Ref. \cite{baghdasaryan2022generalized}), the plot for the state $\ket{\psi_3}$ pumped by $\ell_{\pump}=-1$ is analogous.

Maximal fidelity is possible with extreme waist configurations, either a small pump waist with large collection waists or vice versa, but the choice significantly reduces the pair generation efficiency \cite{coccia2023optimal}. This is because the spatial overlap integral between the pump and down-converted photon modes is diminished. To increase the overlap, one can choose $w_{\pump}$ and $w_{\signal}$ to be close to each other. For example, the yellow curve peaks at $w_{\pump}=15 \mu$m and $w_{\signal}=w_{\idler}=25 \mu$m, yielding a high fidelity of $\mathcal{F}=0.985$. 

For pump OAM modes with $\ell_{\pump} \geq 2$, it is not possible to find equalizable target modes that all have $N_R=0$. Fig.  \ref{fig08} (b) displays the $\ket{1,1}$ and $1/\sqrt{2} \bigl( \ket{1,2}+\ket{2,1} \bigl)$ state pumped with $\ell_{\pump}=2$ and $\ell_{\pump}=3$. Here, the target modes with the highest amplitude are chosen to be the targets. Due to symmetry, for odd $\ell_{\pump}$, those are automatically two modes. The fidelity is fundamentally bounded to a value $\mathcal{F} <1$ due to the presence of conserved but unintended modes with a RMN of $N_R=0$. These modes have fixed amplitude ratios, like $P^{0,2}/ P^{1,2}=1/2$ or $P^{0,3}/ P^{1,2}=1/3$ and cannot be tuned by adjusting the waists. The undesired presence leads to an unavoidable reduction in the overall fidelity of the desired state.

\subsection{Preparation of maximally entangled ququarts}

Previous proposals for creating entanglement by path identity have primarily considered crystals pumped by a Gaussian mode ($\ell_{\pump}=0$). The separable $\ket{\psi_1}$ state yields a high pair generation rate \cite{franke2002two,kysela2020path}, but limits the achievable entanglement dimensionality. If the state $ \ket{\psi_2}$ or $\ket{\psi_3}$ is generated in each crystal, a higher dimensionality $d=2n$ of the entangled state can be achieved compared to $\ket{\psi_1},$ for given number of crystals $n$.

Here, we will discuss two-crystal setups where the crystals are pumped by an $\ell_{\pump}=1$ beam preparing the high-fidelity Bell state $\ket{\psi_2}$ in each crystal. Due to the property $C^{\ell_{\signal},\ell_{\idler}}=(C^{-\ell_{\signal},-\ell_{\idler}})^*$, analogous arguments hold for the state $\ket{\psi_3}$ pumped by $\ell_{\pump}=-1$ (see Appendix \ref{Appendix: Fidelity bounds reason}).

When the paths of two crystals, each pumped by $\ell_{\pump}=1$, are coherently overlapped, entangled ququarts ($d=4$) can be prepared, representing the optimal entanglement dimensionality for a two-crystal path identity configuration. The state will have the general form of
\begin{widetext}

\begin{align}
   \ket{\Psi} = \frac{1}{2} \Biggl( &\mathrm{e}^{i(\varphi_1+\varphi_2)} \ket{\Delta \ell_{A_1}+\Delta \ell_{A_2}, 1+ \Delta \ell_{B_1}+\Delta \ell_{B_2}} +  \mathrm{e}^{i (\varphi_1+\varphi_2)} \ket{1+\Delta \ell_{A_1}+\Delta \ell_{A_2},\Delta \ell_{B_1}+\Delta \ell_{B_2}} \nonumber \\
   + \; & \mathrm{e}^{i \varphi_2} \ket{\Delta \ell_{A_2},1+\Delta \ell_{B_2}} +  \mathrm{e}^{i \varphi_2} \ket{1+\Delta \ell_{A_2},\Delta \ell_{B_2}} \Biggl),
\end{align}

\end{widetext}
and the general setup is shown in Fig. \ref{fig09}. Depending on how the four OAM mode shifts are chosen, we can design MESs for $d=4$. The condition $|\Delta \ell_{{A,B}_1}+\Delta \ell_{{A,B}_2}| + |\Delta \ell_{{A,B}_2}| \geq 2$ is necessary to guarantee that the OAM modes are not partially separable. Shifting the OAM of the photon pair from crystal $\mathsf{1}$ far enough relative to the pair from crystal $\mathsf{2}$ prevents any separability of modes in the superposition.

\begin{figure}[b]
    \centering
    \includegraphics[width=0.9\linewidth]{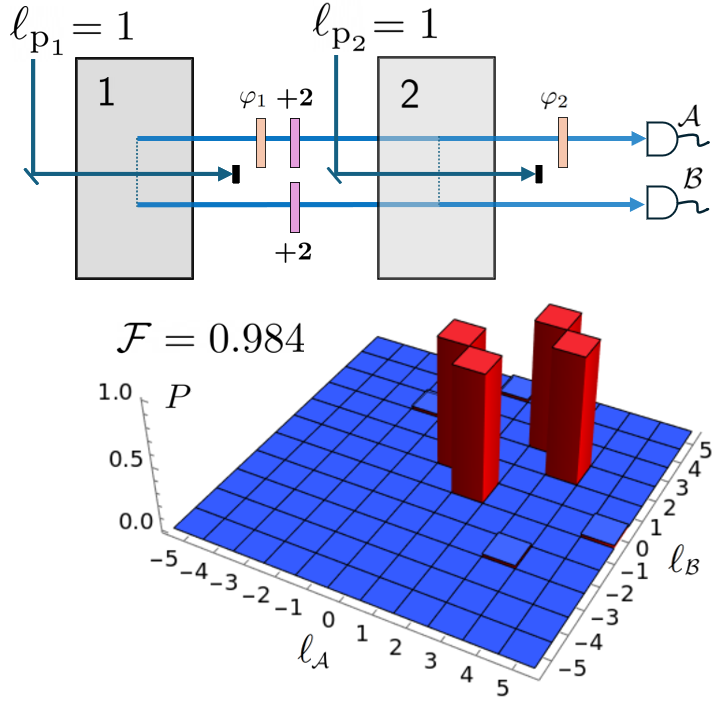}
    \caption{Setup for the generation of the four-dimensional MES $\ket{\psi_{\mathrm{tar}}} = 1/2 \Bigl( \mathrm{e}^{i \varphi_2} \ket{0,1} +  \mathrm{e}^{i \varphi_2} \ket{1,0} +  \mathrm{e}^{i(\varphi_1+\varphi_2)} \ket{2,3} +  \mathrm{e}^{i (\varphi_1+\varphi_2)} \ket{3,2} \Bigl)$ by path identity, along with the resulting OAM spectrum. The fidelity is near unity, meaning all OAM modes are fully concentrated on the target state.} 
    \label{fig010}
\end{figure}

\begin{figure}[b]
    \centering
    \includegraphics[width=.9\linewidth]{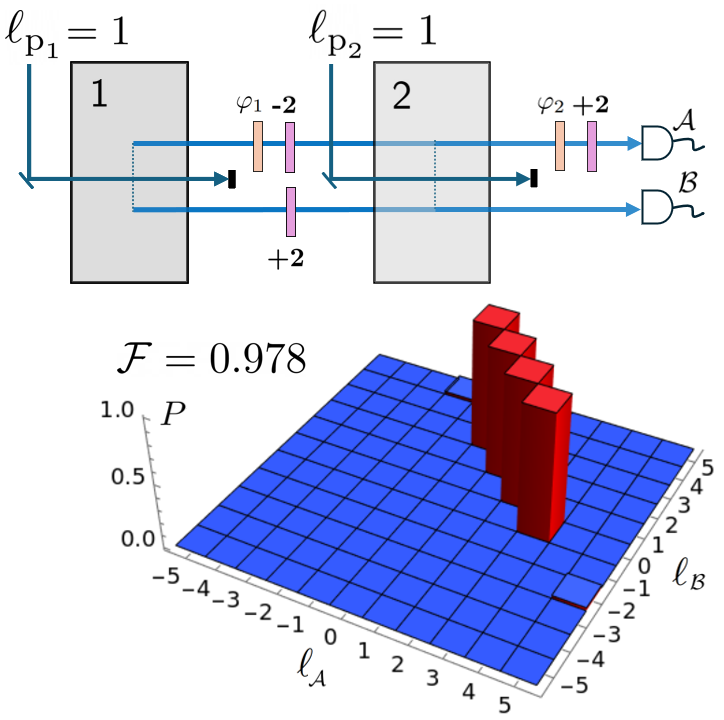}
    \caption{Alternative design for an path identity setup to generate the four-dimensional MES $\ket{\psi_{\mathrm{tar}}} = \frac{1}{2} \Bigl( \mathrm{e}^{i (\varphi_1 +\varphi_2)} \ket{0,3} + \mathrm{e}^{i\mathrm{e}^{i (\varphi_1 +\varphi_2)} } \ket{1,2} + \mathrm{e}^{i \varphi_2} \ket{2,1} + \mathrm{e}^{i \varphi_2} \ket{3,0} \Bigl)$ and the corresponding OAM spectrum. The fidelity is near unity, meaning all OAM modes are fully concentrated on the target state.} 
    \label{fig011}
\end{figure}

We will now construct high-fidelity MES in the OAM space $\ell_{\mathcal{A}}=\ell_{\mathcal{B}}= \bigl \{ \ket{0}, \ket{1},\ket{2},\ket{3} \bigl \}$. A possible setup is proposed in Fig. \ref{fig010}. If the photon pair is created in crystal $\mathsf{1}$, the OAM is shifted by a $\Delta \ell_{A_1}=\Delta \ell_{B_1}=2$ operation. The pair generated in crystal $\mathsf{2}$ is left unchanged. The resulting superposition corresponds to two MESs:
\begin{equation*}
    \ket{\psi_{\mathrm{tar}}} = \frac{1}{2} \Bigl( \ket{0,1} + \ket{1,0} \pm \ket{2,3} \pm \ket{3,2} \Bigr),
\end{equation*}
if the phases are set to $\varphi_1 = 0$ or $\varphi_1 = \pi$, respectively, while $\varphi_2 = 0$. Similarly, Fig. \ref{fig011} shows the generation setup for the MESs given by
\begin{equation*}
    \ket{\psi_{\mathrm{tar}}} = \frac{1}{2} \Bigl( \pm \ket{0,3} \pm  \ket{1,2} + \ket{2,1} + \ket{3,0} \Bigr),
\end{equation*}
when $\varphi_2 = 0$ and $\varphi_1 = 0$ or $\varphi_1 = \pi$, respectively. A crucial aspect of this setup is that the OAM shifts along paths $\mathcal{A}$ and $\mathcal{B}$ have opposite signs. A consecutive setup might use q-plates, which shift the OAM differently for photons with orthogonal polarization, as is the case in type-II SPDC.

These states are a generalization of the Bell state $\ket{\psi_2}$ to $d=4$ and valuable for various applications in quantum information processing. As we cannot separate the entangled OAM modes prepared in a single crystal, we can only manipulate them as a whole and retain the original intrinsic structure of $\ket{\psi_2}$. A major advantage of our setups is that only the paths from two crystals need to be overlapped.

In contrast, to generate a HDES state like 
\begin{equation}
    \ket{\psi_{\mathrm{tar}}} = \frac{1}{2} \Bigl( \ket{0,0} + \ket{1,1} + \ket{2,2} + \ket{3,3} \Bigl),
    \label{state00112233}
\end{equation}
we need to design a four-crystal setup, where each crystal prepares a separable $\ket{\psi_1}$ state, posing greater experimental demands in terms of setup complexity and coherence conditions. Here, each term from Eq. \eqref{state00112233} is produced by a crystal and an appropriate OAM mode shift. 

The transition from a separable state ($d=1$) to an entangled Bell state ($d=2$) in each crystal effectively doubles the dimensionality $d_{\mathrm{PI}}$ for the HDES via path identity.
As a result, we can halve the number of crystals needed to achieve a given total dimensionality. This also means that setups using an equal number of crystals pumped by $\ell_{\pump}=1$ instead of $\ell_{\pump}=0$ can yield a HDES with twice the dimension $2d_{\mathrm{PI}}$ .

A better scaling is closely tied to engineering high-fidelity entangled states with dimensionality $d>2$ from a single crystal. A potential solution to this challenge and overcoming the fidelity bounds presented may lie beyond tuning the pump beam, as a customized phase-matching function \cite{bernecker24high} has been shown to also affect the amplitude of OAM modes with RMNs $N_R<0$. 

\section{Conclusion}

This paper provides an intuitive introduction to how path identity interference can be used to generate and engineer HDES using OAM of light. We place special emphasis on achieving states with a high fidelity in the full OAM space, ensuring that projecting into a desired OAM subspace includes nearly all of the photon pairs produced from parametric down-conversion.

We have demonstrated the existence of fundamental fidelity bounds below $\mathcal{F}=1$ for biphoton states generated by OAM pump modes with $|\ell_{\pump}| > 1$. These limitations cannot be overcome by pump-beam shaping or a suitable choice of collection waists. As a consequence, we recommend using path-identity setups where individual crystals are configured to prepare high-fidelity building blocks, specifically $\ket{0,0}$ (from $\ell_{\pump}=0$) or $1/\sqrt{2} \Bigl( \ket{0,\pm1} + \ket{\pm1,0} \Bigl)$ (from $\ell_{\pump}=\pm1$). 

When using OAM pump modes with $|\ell_{\pump}| > 1$ in a multi-crystal path-identity setup, unintended modes are inevitably generated, which reduces the overall fidelity. These resulting OAM distributions can be realized in a less demanding setup by using a single SPDC process with a spatially engineered pump consisting of an unequally weighted superposition of OAM modes.

Finally, we highlight that a path-identity setup with two crystals, each pumped by an OAM mode of $\ell_{\pump}=1$, can generate a $d=4$ MES. We found this to be the optimal dimensionality, as any attempt to increase the dimensionality of the HDES further results in a fidelity below  $\mathcal{F}<1$. Our proposed configuration allows for a reduction in the number of required crystals by half compared to equivalent states produced by setups with four crystals pumped by Gaussian beams.

Path identity also appears promising for generating multiphoton states, such as GHZ states, using either bulk \cite{wang2024entangling} or integrated optics \cite{bao2023very, hu2025observation}. An interesting direction for future research would be to investigate whether the complexity of such setups could be reduced by using crystals that emit entangled states, rather than separable states, similar to the bipartite case demonstrated in our work.

\begin{acknowledgments}
The authors thank R. Sondenheimer and F. Steinlechner for useful discussions. This research was supported by funding from the Research School of Advanced Photon Science (RS-APS) at the Helmholtz Institute Jena, Germany.
\end{acknowledgments}

\appendix

\section{Fidelity bounds reason}
\label{Appendix: Fidelity bounds reason}
In this section we provide some more detail to calculate the expansion amplitudes from Eq. \eqref{OverlapAmplitudes}. The full expression for the LG modes is
\begin{eqnarray}
     \mathrm{LG}^{\ell}_{p}(\bm{q},w) &&= \sqrt{\frac{w^2 p!}{2\pi(p + |\ell|)!}} (-1)^{p+\frac{\ell}{2}} \left( \frac{|\bm{q}|}{\sqrt{2} w} \right)^{|\ell|} \nonumber \\ && \times \mathrm{e}^{-\frac{|\bm{q}|^2 w^2}{4}} \; \mathrm{L}_{p}^{|\ell |}\left(\frac{|\bm{q}|^2 w^2}{2}\right) \mathrm{e}^{i \ell \mathrm{Arg}(\bm{q}) }.
\label{LGexpression}
\end{eqnarray}
Here, $\mathrm{L}_{p}^{|\ell|} \left( \cdot \right)$ are the generalized Laguerre polynomials, and $w$ is the beam waist. LG modes are characterized by two mode numbers. The radial index $p$ is a positive integer that refers to the nodes observed in the transverse plane. The spiral index $\ell \in \mathbb{Z}$ accounts for the twists in the phase, which result in a projection of $\ell \hbar$ OAM onto the $z$-axis.  We assume the beams are focused on the center of the nonlinear crystal. Inserting Eq. \eqref{LGexpression} into Eq. \eqref{GeneralBiphotonstate}, we arrive at
\begin{widetext}
\begin{small}
\begin{equation*}
    C^{\ell_{\signal},\ell_{\idler}}_{p_{\signal},p_{\idler}} \bigl( \underbrace{p_{\pump},\ell_{\pump},w_{\pump}}_{\mathrm{pump}}; \underbrace{\chi^{(2)},L}_{\mathrm{crystal}}; \underbrace{w_{\signal},w_{\idler}}_{\mathrm{collection}} \bigl) = \; \mathcal{N} \int_{-\frac{L}{2}}^{\frac{L}{2}} dz \iint d\bm{q}_{\signal} \; d\bm{q}_{\idler} \; \;  \chi^{(2)}(z)  \; \mathrm{e}^{ i \Delta k_z(\bm{q}_{\signal},\bm{q}_{\idler}) z} \; \mathrm{LG}_{p_{\mathrm{p}}}^{\ell_{\pump}}(\bm{q}_{\signal}+\bm{q}_{\idler}, w_{\pump}) \left[ \mathrm{LG}_{p_{\signal}}^{\ell_{\signal}}(\bm{q}_{\signal},w_{\signal}) \right]^* \left[ \mathrm{LG}_{p_{\idler}}^{\ell_{\idler}} (\bm{q}_{\idler},w_{\idler}) \right]^*.
\end{equation*}
\end{small}
\end{widetext}

When assuming periodic poling structure for $\chi^{(2)}(z)$ \cite{dosseva2016shaping} such that for the center frequencies $k_{\pump}-k_{\signal}-k_{\idler}=0$, that leaves the mismatch at 
\begin{equation*}
    k_z(\bm{q},\omega) = \frac{| \bm{q}_{\signal}|^2 }{ 2k_{\signal}(\omega_{\signal})}  + \frac{ |\bm{q}_{\idler}|^2 }{ 2k_{\idler}(\omega_{\idler})} - \frac{|\bm{q}_{\signal}+\bm{q}_{\idler}|^2 }{ 2k_{\pump}(\omega_{\pump})}.
\end{equation*}
Further the LG modes from Eq. \ref{LGexpression} can be decomposed via \cite{baghdasaryan2022generalized}
\begin{equation}
         \mathrm{LG}^{\ell}_{p}(\bm{q},w) = \mathrm{e}^{-\frac{|\bm{q}|^2 w^2}{4}} \; \mathrm{e}^{i \ell \mathrm{Arg}(\bm{q}) } \sum_{u=0}^p T_u^{p,\ell}(w) \; | \bm{q}|^{2u + |\ell|}
         \label{LGwithsum}
\end{equation}
where
\begin{align}
   T_u^{p,\ell}(w) &=  \sqrt{\frac{p!\,(p+|\ell|)!}{\pi}} \;  \biggr(\frac{ w}{\sqrt{2}}\biggl)^{2u+|\ell|+1} \nonumber \\
    & \times \frac{(-1)^{p+u} \; i^{\ell}}{(p - u)! \; (|\ell| + u)! \; u!} .
    \label{Tfactor}
\end{align}
Inserting Eqs. \eqref{LGwithsum} and \eqref{Tfactor} into the expression for the coincidence amplitude, we can switch to polar coordinates and simplify the expression analytically, as done very detailed in Ref. \cite{baghdasaryan2022generalized}. 

As described in the main text, the conditions $k_{\signal}=k_{\idler}$ and $w_{\signal}=w_{\idler}$ imply equal expansion amplitudes for the generated OAM modes in SPDC, i.e., $C^{\ell_{\signal}, \ell_{\idler}} = C^{\ell_{\idler}, \ell_{\signal}} $. Under these assumptions, and considering the projection $p=p_{\signal}=p_{\idler}=0$, the full expression can be written as two parts
\begin{equation}
    C^{\ell_{\signal},\ell_{\idler}} \propto T^{\ell_{\pump}, \ell_{\signal}, \ell_{\idler}}(w_{\pump},w_{\signal},w_{\idler}) \cdot \int_{-\frac{L}{2}}^{\frac{L}{2}} dz \; G^{\ell_{\signal},\ell_{\idler}}(z).
    \label{CinAppendix}
\end{equation} 
Here, we define
\begin{align}
    & T^{\ell_{\pump}, \ell_{\signal}, \ell_{\idler}} (w_{\pump},w_{\signal},w_{\idler})  \nonumber \\
    &
    = T_0^{0,\ell_{\pump}}(w_{\pump}) \cdot T_0^{0,\ell_{\signal}}(w_{\signal}) \cdot T_0^{0,\ell_{\idler}}(w_{\idler}) \nonumber \\
    &\propto w_{\pump}^{|\ell_{\pump}|+1} \; w_{\signal}^{|\ell_{\signal}|+1} \; w_{\idler}^{|\ell_{\idler}|+1}.
    \label{Tscaling}
\end{align} 
The kernel in the integral is a cumbersome function of $z$ given by
\begin{align*}
G^{\ell_{\signal},\ell_{\idler}}(z) &\propto \sum_{n=0}^{\ell_{\pump}} \binom{\ell_{\pump}}{n} \Gamma \bigl[b \bigl] \Gamma\bigl[ h(n) \bigl] \frac{D(z)^{d(n)}}{B(z)^{b+h(n)}} \; \\
& \times \: _2\tilde{F}_1 \Bigl[h(n),b,1+d(n),\frac{D(z)^2}{B(z)^2}\Bigl]
\end{align*} 
with the abbreviations
\begin{eqnarray*}
D(z) &&= -\frac{w_{\pump}^2}{4} - \frac{iz}{2 k_{\pump}}, \\
B(z) &&= \frac{w_{\pump}^2}{4} + \frac{w_{\signal}^2}{4}  - iz \frac{k_{\pump}-k_{\signal,\idler}}{2 k_{\pump}k_{\signal}}, \\
b&&=1+\frac{\ell_{\idler}}{2}+\frac{|\ell_{\idler}|}{2}, \\
h(n)&&=1+\frac{\ell_{\pump}}{2}+\frac{\ell_{\idler}}{2}+\frac{|\ell_{\signal}|}{2}-n, \\
d(n)&&=\ell_{\idler} - n.
\end{eqnarray*}
The function $G^{\ell_{\signal},\ell_{\idler}}(z)$ can actually be distinguished in two cases,
\begin{widetext}
\begin{align*}
G^{\ell_{\signal},\ell_{\idler}}(z) = & 
\begin{cases}
\displaystyle
\frac{(\ell_{\signal} + \ell_{\idler})!}{
\Bigl( B(z) - D(z))^{\ell_{\signal} + \ell_{\idler} + 1} \cdot (B(z) + D(z) \Bigl)
} & \mathrm{if \;} \ell_{\signal} \geq 0 \mathrm{, } \ell_{\idler} \geq 0 \\[12pt]
\displaystyle
\frac{\max \Bigl( |\ell_{\signal}|, |\ell_{\idler}| \Bigl)! \; \; D(z)^{\min\left( |\ell_{\signal}|, |\ell_{\idler}| \right)}
}{\Bigl(B(z) - D(z) \Bigl)^{\max\bigl( |\ell_{\signal}|, |\ell_{\idler}| \bigl) + 1} \; \; \bigl( B(z) + D(z) \bigl)^{\min\bigl( |\ell_{\signal}|, |\ell_{\idler}| \bigl) + 1}
} & \mathrm{if \;} \ell_{\signal} < 0,\ell_{\idler} \geq 0  \mathrm{\; or \;}  \ell_{\signal} \geq 0,\ell_{\idler} < 0
\end{cases}
\end{align*}
\end{widetext}

\begin{figure}[t]
    \centering
    \includegraphics[width=0.95\linewidth]{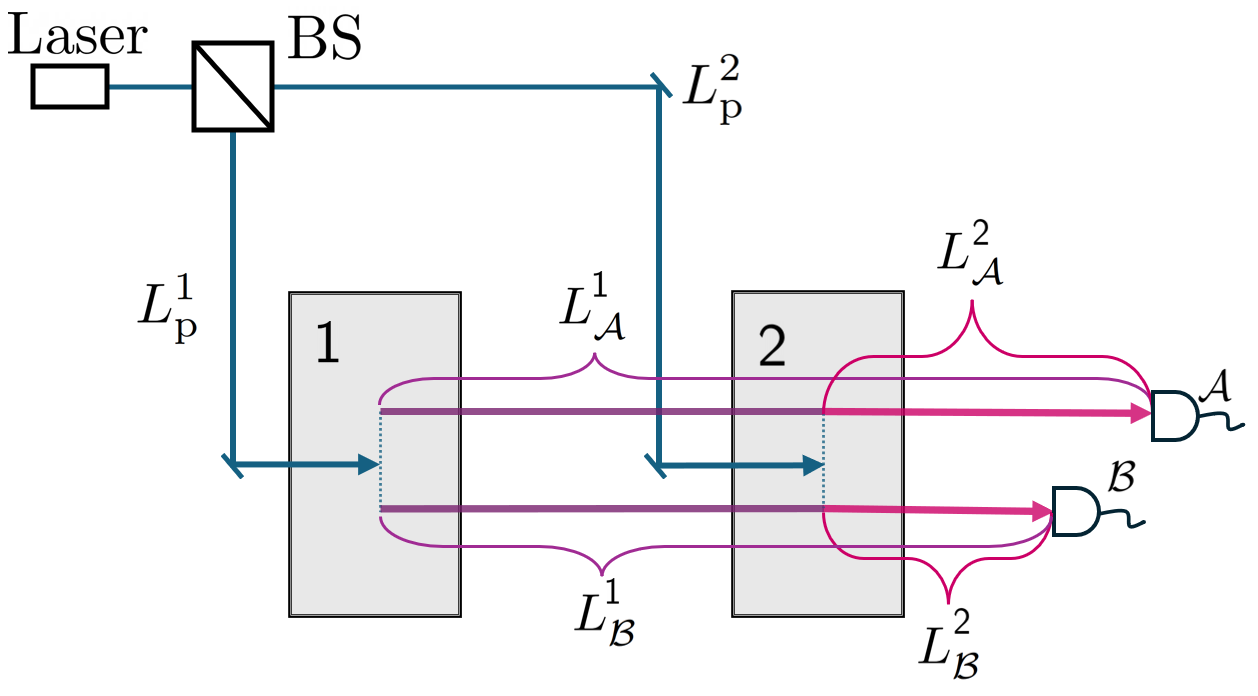}
    \caption{Path-identity setting for two crystals. A laser produces a pump beam, which is split by a beam splitter (BS) to coherently pump crystals $\mathsf{1}$ and $\mathsf{2}$. The generation of a pair is triggered in either crystal and detected at detectors along $\mathcal{A}$ and $\mathcal{B}$.}
    \label{fig012}
\end{figure}

This explains the special behavior for $\ell_{\signal} \geq 0 \mathrm{, } \ell_{\idler} \geq 0$. 
For all coincidence amplitudes fulfilling the condition, the integral in Eq. \eqref{CinAppendix} give the same result. When looking at  \eqref{Tscaling}, the dependence of the beam waists for $w_{\signal}=w_{\idler}$ will cancel out for the ratio $\frac{C^{\ell_{\signal_1},\ell_{\idler_1}} }{C^{\ell_{\signal_2} \ell_{\idler_2}} }$ if $\ell_{\signal},\ell_{\idler}>0$, . We are left with the ratio of the proportional constant  from Eq. \eqref{Tscaling}, which is given by
\begin{equation*}
  \frac{C^{\ell_{\signal_1},\ell_{\idler_1}} }{C^{\ell_{\signal_2},\ell_{\idler_2}} } =
  \sqrt{\frac{\ell_{\signal_2}! \; \ell_{\idler_2}! }{\ell_{\signal_1}! \; \ell_{\idler_1}!}}
\end{equation*}
for amplitudes pumped by the same OAM mode, i.e., $\ell_{\pump}= \ell_{\signal_1}+\ell_{\idler_1}=\ell_{\signal_2}+\ell_{\idler_2}$
In conclusion, for $\ell_{\signal},\ell_{\idler}>0$ the ratios of different modes and therefore their projection probability is fixed. The ratio between the projection probabilities results in
\begin{equation*}
  \frac{P^{\ell_{\signal_1},\ell_{\idler_1}} }{P^{\ell_{\signal_2},\ell_{\idler_2}} } = \frac{\ell_{\signal_2}! \; \ell_{\idler_2}! }{\ell_{\signal_1}! \; \ell_{\idler_1}!}.
\end{equation*}
Since for $\ell_{\pump} = \ell_{\signal}+ \ell_{\idler} <0$ we use $C^{\ell_{\signal}, \ell_{\idler}}=\bigl( C^{-\ell_{\signal},- \ell_{\idler}} \bigl)^*$, the same argument holds for the region $\ell_{\signal} \leq 0 \mathrm{, } \ell_{\idler} \leq 0$. 

\section{Temporal Coherence Conditions}
\label{tempindist}

\begin{figure*}
    \centering
    \includegraphics[width=0.85\linewidth]{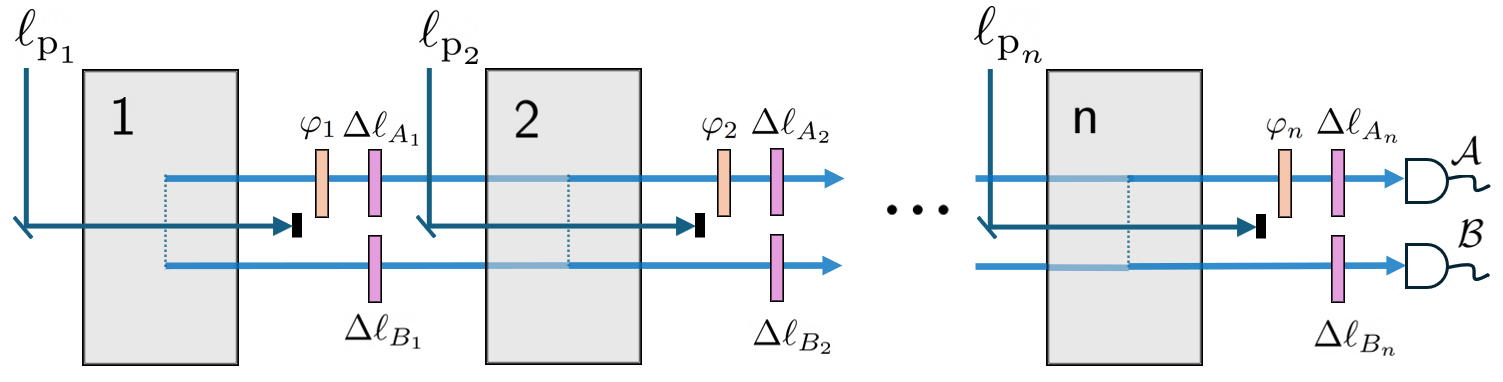}
    \caption{The modular two-crystal path identity configuration can be easily generalized to a setup with $n$ crystals. Here, photon pairs from crystal $\mathsf{1}$ undergo $n$ mode and phase shifts in total, given by $\Delta \ell_{A_1}+\Delta \ell_{A_2}+\dots+ \Delta\ell_{A_n}, \Delta \ell_{B_1}+\Delta \ell_{B_2}+\dots+\Delta \ell_{B_n}$ and $\varphi_1+\varphi_2+\dots+\varphi_n$; photon pairs from crystal $\mathsf{2}$ undergo $n-1$ mode and phase shifts in total, $\Delta\ell_{A_2}+\dots+ \Delta \ell_{A_n}, \Delta \ell_{B_2}+\dots+\Delta \ell_{B_n}$ and $\varphi_2+\dots+\varphi_n$; and so on. Photon pairs from crystal $\mathsf{n}$ undergo a single mode and phase shift $\Delta \ell_{A_n}, \Delta \ell_{B_n}$ and $
    \varphi_n$. The mode shifters may transform the OAM of the photons to values that are not available solely from the down-conversion process. This allows enlarging the OAM Hilbert space. Different pumping strengths of OAM pump modes $\ell_{p_1}, \dots, \ell_{p_n}$ can be used as new tuning parameters to increase or decrease the probability of pairs being created in crystals $\mathsf{1}, \dots, \mathsf{n}$.}
    \label{fig013}
\end{figure*}

Fig.  \ref{fig012} shows an interferometric setup, which is a common implementation for coherently pumping two crystals. A single pump beam is split into two parts, which are then individually transformed into modes with OAM  $\ell_{p_1}$ and $\ell_{p_2}$. To ensure temporal indistinguishability, the following two coherence conditions are crucial to fulfill. Firstly, we must ensure that we cannot distinguish between the arrival times of the signal and idler photons, which is guaranteed by
\begin{equation}
|L_{\mathcal{A}}^{j} - L_{\mathcal{B}}^{j}| < L^{j}_{\mathrm{coh, DC}}
\end{equation}
for $j=\mathsf{1,2}$. Here, $L_{\mathcal{A,B}}^{j}$ is the optical path length of the photons from crystal $\mathsf{j}$ along $\mathcal{A}$ or $\mathcal{B}$ and $L^{j}_{\mathrm{coh, DC}}$ the coherence time of the down-converted pair. Secondly, we must not have temporal information that reveals whether the photon pairs were generated in crystal $\mathsf{1}$ or $\mathsf{2}$, which is satisfied by
\begin{equation*}
|L_{\pump}^1 - L_{\pump}^2 - L_{\mathrm{DC}}| < L_{\mathrm{coh, p}}.
\end{equation*}

Here, $L_{\mathrm{coh, p}}$ is the coherence length of the pump, $L_{\pump}^{1,2}$ are the optical path lengths from the splitting point to the crystals $\mathsf{1,2}$ and $L_{\mathrm{DC}}$ is the path length of the down-converted photons. Note that $L_{\mathrm{DC}} \leq L_{\mathcal{A}}^{\mathsf{1}}$. Experimentally, the indistinguishability can be tested via visibility measurement when detecting simultaneous two-fold coincidences when the optical path length is altered. A perfect visibility of $V=1$ indicates that the crystals are perfectly indistinguishable \cite{hochrainer2022quantum}.

\section{Generalized scheme}
\label{n crystal scheme}

Here, we generalize the two-crystal path identity setup discussed in the main paper to a setup involving $n$ crystals, labeled $\mathsf{1}$ to $\mathsf{n}$. Eq. \eqref{PIEtwocrystals} transforms to
\begin{widetext}
\begin{align*}
\ket{\Psi} &=  \frac{1}{\sqrt{N}} \Biggl(\mathcal{X}_1 \sum_{\ell_1=-\infty}^{\infty} C^{\ell_1,\ell_{p_1}-\ell_1} \; \mathrm{e}^{i ( \varphi_1 + \varphi_2 + ... +\varphi_n )} \; \ket{\ell_1 + \Delta \ell_{A_1}+ \Delta \ell_{A_2}+...+\Delta \ell_{A_n}}_{\mathcal{A}} \otimes \ket{\ell_{p_1}-\ell_1+ \Delta \ell_{B_1} + \Delta \ell_{B_2}...+ \Delta \ell_{B_n}}_{\mathcal{B}} \nonumber \\
&+ \mathcal{X}_2 \cdot  \sum_{\ell_2=-\infty}^{\infty} C^{\ell_2,\ell_{p_2}-\ell_2} \; \mathrm{e}^{i ( \varphi_2+ ... +\varphi_n)} \; \ket{\ell_2 + \Delta \ell_{A_2}+...+ \Delta \ell_{A_n}}_{\mathcal{A}} \otimes \ket{\ell_{p_2}-\ell_2+ \Delta \ell_{B_2}+...+\Delta \ell_{B_n}}_{\mathcal{B}} \nonumber \\
& ... \nonumber \\
&+ \mathcal{X}_n \cdot  \sum_{\ell_n=-\infty}^{\infty} C^{\ell_n,\ell_{p_n}-\ell_n} \; \mathrm{e}^{i \varphi_n} \; \ket{\ell_n + \Delta \ell_{A_n}}_{\mathcal{A}} \otimes \ket{\ell_{p_n}-\ell_n + \Delta \ell_{B_n}}_{\mathcal{B}} \Biggl)
\label{PIEncrystals}
\end{align*}
\end{widetext}
where we placed $n$ OAM shifters along path $\mathcal{A}$, $\Delta \ell_{A_{1}...,n}$ and $\mathcal{B}$, $\Delta \ell_{B_{1}...,n}$, respectively, as well as $n$ phase shifters $\varphi_{1,...,n}$. Fig. \ref{fig013} sketches this scenario. $\mathcal{X}_1, \mathcal{X}_2,...,\mathcal{X}_n$ are the relative pump powers. In this case, the updated (re-)normalization constant reads
\begin{widetext}
\begin{equation*}
    N = |\mathcal{X}_1|^2 \cdot \sum_{\ell_1} |C^{\ell_1,\ell_{p_1}-\ell_1}|^2 +  |\mathcal{X}_2|^2 \cdot  \sum_{\ell_2} |C^{\ell_2,\ell_{p_2}-\ell_2}|^2 + ... +  |\mathcal{X}_n|^2 \cdot  \sum_{\ell_n} |C^{\ell_n, \ell_{p_n}-\ell_n}|^2.
\end{equation*}
\end{widetext}
With $n$ stages of mode shifters, we retain the flexibility to generate states that require specific shifts within the desired Hilbert space. However, setups are not unique. For instance, when a crystal emits a mode that is already part of the final superposition, we can assign this crystal as number $n$, and only $n-1$ mode shifters are needed. This is the case for the state shown in Fig. \ref{fig07}, which can be equivalently realized with three stages of mode shifters as $\Delta\ell_{A_1} = \Delta\ell_{B_1} = -2$, $\Delta\ell_{A_2} = \Delta\ell_{B_2} = -2$, and $\Delta\ell_{A_3} = \Delta\ell_{B_3} = 2$. In Fig.~11, for example, the mode shifter after the last crystal is essential to obtain the desired MES.

\bibliography{apssamp}

\end{document}